\documentclass[prd,nofootinbib,showpacs,superscriptaddress]{revtex4}

\usepackage{bm}
\usepackage{amsfonts}

\newcommand{\mb}[1]{\mbox{\boldmath $#1$}}
\def \be {\begin{equation}}
\def \ee {\end{equation}}
\def \beq {\begin{eqnarray}}
\def \eeq {\end{eqnarray}}
\def \bean {\begin{eqnarray*}}
\def \eean {\end{eqnarray*}}
\def \d   {\delta}
\def \D   {{\Delta}}

\def \la  {\lambda}
\def \ga  {\gamma}
\def \om  {\omega}

\def \nn  {\nonumber}
\def \met {\mbox{g}}
\def \metb{\mbox{\bf g}}

\def \part {{\partial}}

\begin{document}

\title[Coupling of radial and non-radial oscillations of relativistic stars]
{Coupling of radial and non-radial oscillations of relativistic stars: \\
gauge-invariant formalism}

%
\author{Andrea Passamonti}
\affiliation{Institute of Cosmology and Gravitation,
University of Portsmouth, Portsmouth PO1 2EG, United Kingdom}
\author{Marco Bruni}
\affiliation{Institute of Cosmology and Gravitation,
University of Portsmouth, Portsmouth PO1 2EG, United Kingdom}
\author{Leonardo Gualtieri}
\affiliation{Dipartimento di Fisica ``G. Marconi'', Universit\`a di Roma
``La Sapienza'' and Sezione INFN ROMA 1, piazzale Aldo Moro 2, I--00185 Roma,
Italy}
\author{Carlos F. Sopuerta}
\affiliation{Institute of Cosmology and Gravitation,
University of Portsmouth, Portsmouth PO1 2EG, United Kingdom}
\affiliation{Institute for Gravitational Physics and Geometry and
Center for Gravitational Wave Physics, \\
Penn State University, University Park, PA 16802, USA}


\date{\today}

\begin{abstract}

Linear perturbation theory is appropriate to describe  small
oscillations of stars, while a mild non-linearity is still tractable perturbatively but
requires to consider mode coupling, i.e.\ to take into account second order
effects.
It is natural to start to look at this problem by considering the coupling
between linear radial and non-radial modes.  A radial pulsation may be
thought of as an important  component of an overall mildly non-linear
oscillation, e.g.\ of a proto-neutron star.
Radial pulsations of a spherical compact objects do not {\it per se} emit
gravitational waves but,
if the coupling between the existing first order radial and non-radial modes is efficient in driving and possibly amplifying the non-radial oscillations, one may expect the appearance of non-linear harmonics, and  gravitational radiation could then be produced to a significant
level. More in general, mode coupling typically  leads to an interesting
phenomenology, thus it is worth investigating it in the context of star
perturbations.

In this paper we  develop the relativistic formalism to study  the coupling
of radial and non-radial
first order  perturbations of a compact spherical star.  From a mathematical
point of view, it is convenient to treat the two sets
of perturbations as separately parametrized,
using a 2-parameter perturbative expansion of the metric, the
energy-momentum tensor and  Einstein
equations in which $\lambda$ is associated with the radial modes,
$\epsilon$ with the non-radial perturbations, and the $\lambda\epsilon$ terms
describe the coupling. This approach provides
 a well-defined framework to   consider the
gauge dependence of perturbations,  allowing  us to use  $\epsilon$ order gauge-invariant
non-radial variables on the static background and to define new
second order $\lambda\epsilon$ gauge-invariant variables representing the result of the
non-linear coupling.  We present  the evolution and constraint equations
for our variables outlining the setup for numerical computations, 
and briefly discuss the surface boundary
conditions in terms of the second order $\lambda\epsilon$ Lagrangian
pressure perturbation.
\end{abstract}


\maketitle

\section{Introduction}

Since the discovery of pulsars in the sixties neutron stars  have
acquired a special status in physics: as supernovae remnants they are
fundamental to our understanding of the final stages of evolution and
fate of upper main sequence stars; as the most compact observed
objects they are a test-bed for strong-field gravity, i.e. general relativity  and its
generalizations; they are a unique laboratory for fundamental physics
such as nuclear interactions, superfluidity, superconductivity. Either
as isolated objects or in binary systems neutron stars are important gravitational
wave sources, and in the near future the analysis of gravitational
radiation from neutron stars will open up a direct window on their interior,
possibly revealing details on the equation of state of nuclear matter,
the dynamics of the crust-mantle interaction and the inner
superfluid/superconducting core. Early stages of neutron star formation during
iron core collapse in supernov\ae \ and other possible transient dynamical phases are
particularly interesting from the point of view of gravitational wave
physics.  In this case the gravitational radiation emitted should carry the
signature of the yet not understood mechanism driving the explosion and that of the
unknown phenomena responsible for the observed kick
velocity of many pulsars, providing complementary information to that
carried by neutrinos.

Physics of compact objects like supernov\ae \ core collapse or neutron stars may be
studied with various approaches and approximations, from purely
Newtonian, or Newtonian with relativistic corrections, to the full general relativistic
treatment of numerical relativity.  For many astrophysical problems even a purely
Newtonian approach may be adequate.  However, in order to study neutron stars as
sources of gravitational waves, a full general relativistic treatment is more
satisfactory  because in this case the gravitational radiation is built-in in the
calculations rather than
calculated {\it a-posteriori} with the quadrupole formula, thus one
can obtain more accurate  results. As for other gravitational wave sources, in the
long term the goal is a full numerical relativity treatment coupled with a
detailed description of the matter
physics. Meanwhile, relativistic perturbation methods remain a valid alternative:
they are computationally far less intensive because, even in dealing with non-linear
effects as we shall be doing here, one has to solve linear partial differential
equations instead of the complete non-linear Einstein equations. Hence, accurate
results can be obtained at a relatively low cost. Moreover, even when numerical
solutions of the complete field equations
are available, analytic perturbative methods often help to shed light on
the physical processes at work,  clarifying  the interpretation of the 
numerical results.

Linear perturbations and instabilities of neutron stars have been studied for
long time~\cite{Andersson:2002ch,Kokkotas:2002ng} but relatively
little is known of non-linear dynamical effects (see e.g.\
\cite{Sperhake:2001xi,2002PhRvD..65b4001S,2002ApJ...571..435M,2002PhRvD..65h4039L,2003ApJ...591.1129A})
and therefore
second order studies may help to understand known problems and even reveal a new
phenomenology.  Non-linear effects are
the rule rather than the exception in phenomena in all branches of
physics, and have to be taken into account for accurate modeling. This
should even be more natural when the mathematical modeling is not
phenomenological, but rooted in a fundamental theory which is {\it
per se}  non-linear, as general relativity  is. When we use perturbation theory,
we generally expect that if the perturbations are very small, then
second order effects should be negligible. On the other hand, while in
order to treat strong non-linear dynamical effects a fully non-perturbative approach is
required, one may expect that much of the interesting physics only
involves a mild non-linearity for which a second order treatment
should be perfectly suited. Furthermore, a proper parametrization of a
problem may even lead to the unexpected result that a second order
treatment appears to provide a good description of the physics even in
a mildly non-linear regime where {\it a priori} one would expect the
perturbative approach to fail. An example  is given by the study of  black hole
collision  in the close limit approximation
\cite{1999bhgr.conf..351P}. With this  in mind, it is reasonable to expect  that a
second order treatment of non-linear oscillations of neutron stars should be
adequate in many circumstances of astrophysical and/or
gravitational interest.

In this paper we develop the formalism to study a specific non-linear effect,
focusing  our investigation on the coupling of radial and non-radial
first order perturbations of a relativistic star. Work is currently under way  to 
apply this formalism and carry out the actual study, and will be presented
elsewhere \cite{Passamonti:2004axial,Passamonti:2005polar}.  The main goal is that of
understanding whether this coupling can lead to new effects such as fluid instabilities
and/or persistence of oscillations and amplifications that
can produce a significant amount of gravitational radiation. Furthermore, this work may 
be a first step toward a more comprehensive study of second order perturbations of
compact stars and mode coupling.

The physical picture we have in mind here is that of a star undergoing a phase of 
overall oscillation or wobbling that we want to describe going beyond the linear 
approximation.
Purely radial modes are going to be  a
natural component of such dynamical phase; on the other hand, even leaving rotational 
effects aside (as we do here) it is hard to think of oscillations that don't have at
least a tiny non-radial component.
 While purely radial and  non-radial second order effects may also eventually become
 important in the non-linear phase, it is thus natural to first investigate the
 coupling of their linear components.
 A specific motivation for this is  that the purely  radial oscillations of a
 spherical star don't emit {\it per se} any gravitational waves  but, if the
coupling we aim to study is efficient in driving and possibly amplifying the 
non-spherical
oscillations, one may expect the appearance of non-linear harmonics, and gravitational radiation could then be produced to a significant level,  provided that at first order both the radial and the non-radial modes are  non-vanishing. 
More in general, mode coupling typically  leads to an interesting phenomenology, 
thus worth investigating. For example,  a specific effect we can easily anticipate is 
that axial modes, decoupled from fluid perturbations at first order, can be sourced
by radial oscillations.
Finally, a further motivation is that  there are a number of studies aiming at investigating if
non-radial oscillations of stars can be excited by external
sources (see e.g.\ \cite{Gualtieri:2001cm,Poisson:1993vp}). Instead, our general idea is to see if the non-radial oscillations can be driven or even amplified through coupling by an internal  radial oscillation, regardless of the presence of an external source. These sort of non-linear processes could occur, for instance, in a proto-neutron star that is still pulsating. A mainly radial pulsation could for example drive the non-radial oscillations, either naturally present, or excited through fall-back accretion.

Our study can also be seen from a slightly different perspective:
we are investigating linear non--radial perturbations of  a radially
oscillating star, where the radial oscillation is also treated perturbatively.  From
this point of view  our problem is mathematically similar to that of studying
linear perturbations of a
slowly rotating star (see e.g.\ \cite{Ruoff:2001fq}) described by the Hartle-Thorne
metric \cite{Hartle:1967ha,Hartle:1968ht,Berti:2004ny}: in the latter case the
spacetime we perturb is itself a special perturbation - stationary and axisymmetric 
(typically up to
second order) - of a spherical static star. Similarly, from this point of view
we consider here perturbations of a spacetime which is itself a spherically symmetric
time-dependent perturbation of a static spherical star.
As we shall see, this point of view turns out to be of  practical value, although it
must be used with care because of the typical gauge issues of relativistic perturbation
theory.

Mathematically is however more satisfactory  to consider radial and
non-radial perturbations as different first order perturbations of a
static spherically symmetric background that couple at second order.
This point of view renders  transparent two crucial aspects of our problem: {\it i)}
the perturbations are defined as fields on a static spherical background; {\it ii)}
the two sets of perturbations are separately parametrized. Thus  a well-defined
framework for our study is provided by  a multi-parameter relativistic perturbation
formalism previously developed~\cite{Bruni:2002sm,Sopuerta:2003rg}.  This allows us
to set up the formalism in a hopefully transparent way, properly considering the
gauge dependence of perturbations. Fixing the gauge for radial
perturbations, we borrow the formalism developed by
Gundlach and Mart\'{\i}n
Garc\'{\i}a~\cite{Gundlach:1999bt,Martin-Garcia:2000ze} (based on that
of Gerlach and Sengupta~\cite{Gerlach:1979rw}; we shall refer to the
GSGM formalism in the following), which gives equations for
gauge-invariant perturbations on a general time--dependent spherical
background.  This then allows us to: {\it i)} have gauge-invariant
non-radial first order variables on our static background and {\it
ii)} to define new second order variables, describing the non-linear
coupling of the the radial and non-radial linear perturbations, that
are also gauge-invariant at second order.  This higher order gauge
invariance, attained by partially fixing the gauge at first order,
is similar to that considered for example in~\cite{Cunningham:1980cp}
and~\cite{Gleiser:1995gx}, although in  our case we deal with a 2-parameter
expansion~\cite{Bruni:2002sm,Sopuerta:2003rg} and we only need to fix the gauge
for radial perturbations. 

At first order most of the fluid perturbations appear in the polar part of the spectrum.
Hence we expect that the effects of the radial non-radial coupling will predominantly
manifest themselves  through polar modes. Therefore in this paper we focus on  polar
perturbations of a perfect fluid star.

The plan of the paper is the following. In Section II we describe
the general 2-parameter perturbative framework we are going to use. In Section III
we briefly recall the GSGM formalism. Using this later, in Section IV we
introduce the radial and non-radial first order perturbations of a
static, spherically symmetric star. Section V  introduces the
second order perturbations that account for the coupling between the
radial and non-radial first order ones; we prove the gauge invariance
of such perturbations, and give the equations they fullfill.
In Section VI we briefly discuss the problem of the boundary conditions, defining
the second order Lagrangian pressure perturbations needed to fix the problem at 
the surface of the star.  Finally in Section VII we draw our conclusions.

The conventions that we follow throughout this work are:
Greek letters are used to denote spacetime indices;
capital Latin letters are used for indices in the time-radial
part of the metric;  lower-case Latin indices are used for
 the spherical sector of the metric.  We use physical
units in which $8\pi G = c = 1$.

\section{Perturbative Framework} \label{framework}

In order to study the coupling of the first order radial and non-radial perturbations
of a static spherically symmetric star it is convenient to use  a 2-parameter perturbative
approach, where the coupling will then appear at higher order.  A natural framework for
the study of this problem is therefore provided by the multi-parameter non-linear
perturbative formalism introduced in~\cite{Bruni:2002sm,Sopuerta:2003rg}.
Generalising well known mathematical ideas at the basis of standard 1-parameter
linear \cite{Wald:1984cw,1974RSPSA.341...49S} and non-linear \cite{Bruni:1996im,Sonego:1998np}
perturbation theory, the basic underlying assumption in the construction of a multi-parameter
relativistic non-linear formalism is the existence of a multi-parameter
family of spacetime models  that can be Taylor expanded around a {\em background} spacetime,
representing an idealized situation.
These spacetime models are labeled
by a set of parameters that formally control the strength of the perturbations with
respect to the background, and serve as book keeping.  The crucial point
to obtain a manageable theory is to choose a convenient background, in our case the static
spherically symmetric star.

Let us denote the metric of this  background with $\mb{\metb}^{(0,0)}$, i.e.\  a
Tolman-Oppenheimer-Volkov (TOV) solution of the field equations. Denoting with $\mb{\metb}$
the physical metric, we shall expand it in the two parameters $\lambda$ and $\epsilon$, and
we will use superscript indices $(i,j)$ to denote perturbations of order $i$ in  $\lambda$
and $j$ in $\epsilon$.

We then have
\begin{equation}
\met_{\alpha\beta}=\met^{(0,0)}_{\alpha\beta} + \lambda  \,
\met^{(1,0)}_{\alpha\beta}+
\epsilon \, \met^{(0,1)}_{\alpha\beta} + \lambda\epsilon \, \met^{(1,1)}_{\alpha\beta}
+O(\lambda^2,\epsilon^2)\,, \label{initialmetric}
\end{equation}
where the
terms $\mb{\metb}^{(1,0)}$ and $\mb{\metb}^{(0,1)}$ respectively
represent first order radial and non-radial perturbations, and
$\mb{\metb}^{(1,1)}$ is the non-linear  contribution due to
the coupling, which
is the new quantity that we want to compute. The other second order
perturbations, i.e. the self-coupling terms of order $\lambda^2$ and
$\epsilon^2$, will not be considered in this work.
Any  field can be expanded as the metric in Eq.~(\ref{initialmetric}). In particular, we can
expand in this way fluid variables like the energy density and the 4-velocity, and for
the energy-momentum tensor  $\mb{T}$ we can formally write
\begin{equation}
T_{\alpha\beta} = T^{(0,0)}_{\alpha\beta} + \lambda \, T^{(1,0)}_{\alpha\beta}+
\epsilon \,T^{(0,1)}_{\alpha\beta} + \lambda\epsilon \, T^{(1,1)}_{\alpha\beta}
+O(\lambda^2,\epsilon^2) \,, \label{tmunu}
\end{equation}
where each term  $\mb{T}^{(i,j)}$ collects terms of the metric and fluid variables of the appropriate order.
Let us now consider
the structure of the perturbed field  equations, following a standard procedure~\cite{Wald:1984cw}.
We start from thefull Einstein equations:
\begin{equation}
\mb{E}\left[\,\metb\,,\mb{\psi}_A\,\right] = \mb{G}\left[\,\metb\,\right] -
\mb{T}\left[\,\metb\,,\mb{\psi}_A\,\right] = 0 \,, \label{efes}
\end{equation}
where $\mb{G}$ denotes  the
Einstein tensor, and $\mb{\psi}_A$ (A=$1,\dots$)  the various fluid variables.  
If we introduce the perturbative expressions~(\ref{initialmetric}) and (\ref{tmunu})
into  Eq.~(\ref{efes}), we can expand the latter up to $(1,1)$ order,  obtaining
\begin{eqnarray}
&& \! \! \! \! \! \! \mb{E}  \left[\,\metb\,,\mb{\psi}_A\,\right]   =
\mb{E}^{(0,0)}\left[\,\mb{\metb}^{(0,0)}\,,\mb{\psi}_A^{(0,0)}\,\right]
+ \lambda\,\mb{E}^{(1,0)}
\left[\,\mb{\metb}^{(1,0)}\,,\mb{\psi}_A^{(1,0)}\,\right]
+ \epsilon\,\mb{E}^{(0,1)}
\left[\,\mb{\metb}^{(0,1)}\,,\mb{\psi}_A^{(0,1)}\,\right]  \nn
 \\
&&  \! \! \! \! \! \! \! \!
\quad\quad +\lambda\epsilon\,\mb{E}^{(1,1)}\left[\,\mb{\metb}^{(1,1)}\,,\mb{\psi}_A^{(1,1)}~\left|~~
\mb{\metb}^{(1,0)}\otimes\mb{\metb}^{(0,1)}\,, \mb{\psi}_A^{(1,0)}\otimes\mb{\psi}_A^{(0,1)}\,,
\mb{\metb}^{(1,0)}\otimes\mb{\psi}_A^{(0,1)}\,, \mb{\psi}_A^{(1,0)}\otimes\mb{\metb}^{(0,1)} \,\right]
+ O(\lambda^2,\epsilon^2)=0  \,. \right. \label{total}
\end{eqnarray}
The previous
equation is satisfied for arbitrary values of the two parameters if,
and only if, each coefficient of the expansion vanishes.  Therefore, setting each of these
terms to zero, $\mb{E}^{(0,0)}=0$  represents  the TOV equations (see,
e.g.,~\cite{Misner:1973cw}), while  each of the other   $\mb{E}^{(i,j)}=0$ terms represent
the perturbative equations of order $(i,j)$. As differential operators, the $\mb{E}^{(i,j)}$ act
linearly on each of the terms in square brackets, while they are non-linear functions of the
background quantities $\metb^{(0,0)}$ and $\mb{\psi}_A^{(0,0)}$.

At first order in $\lambda$, we obtain the equations describing the radial perturbations
on the TOV  background,
\begin{equation}
\mb{E}^{(1,0)}\left[\,\mb{\metb}^{(1,0)}\,,\mb{\psi}_A^{(1,0)}
\,\right] = 0 \,. \label{linradial}
\end{equation}
Linear radial
perturbations have been extensively analyzed  in the
literature~(see~\cite{Misner:1973cw} and references therein,
and~\cite{Kokkotas:2000up} for more recent results).

The linearized equations for the non-radial perturbations come from the
first order terms in $\epsilon$,
 \begin{equation}
\mb{E}^{(0,1)}\left[\,\mb{\metb}^{(0,1)}\,,\mb{\psi}_A^{(0,1)}\,\right]=0\,, \label{Enonr}
\end{equation}
and  were first studied  by Thorne and Campolattaro~\cite{Thorne:1967th}.  Later they became
the subject of many investigations, see e.g.~\cite{Detweiler:1985dl,Chandrasekhar:1991fi,
Allen:1998xj,Ruoff:2001ux,Seidel:1987in,Seidel:1990xb}.

Finally, the equations describing the radial non-radial coupling, the
ones we shall focus on, have the following form
\begin{equation}
\mb{E}^{(1,1)}\left[\,\mb{\metb}^{(1,1)}\,,\mb{\psi}_A^{(1,1)}~\left|~~
\mb{\metb}^{(1,0)}\otimes\mb{\metb}^{(0,1)}\,, \mb{\psi}_A^{(1,0)}\otimes\mb{\psi}_A^{(0,1)}\,,
\mb{\metb}^{(1,0)}\otimes\mb{\psi}_A^{(0,1)}\,, \mb{\psi}_A^{(1,0)}\otimes\mb{\metb}^{(0,1)} \,\right]
=0\,. \right. \label{EinCoup}
\end{equation}
It is an intrinsic feature of  perturbation theory that  the procedure to solve the above
equations is iterative. Thus,  when we arrive at the stage of solving Eq.~(\ref{EinCoup}), the
terms $\mb{\metb}^{(1,0)}$ and $ \mb{\psi}_A^{(1,0 )}$ are assumed
to be known from solving the radial linear equation (\ref{linradial}), while
$ \mb{\metb}^{(0,1)}$ and $ \mb{\psi}_A^{(0,1 )}$ are solutions of the non-radial linear
perturbation equations (\ref{Enonr}). Hence, because of the linearity of the operator
$\mb{E}^{(1,1)} $ in acting on each of the terms in square brackets, these terms play
the role of sources in Eq.~(\ref{EinCoup}).
Taking into account the nature of the different sets of perturbations we are
considering, it turns out that  the operator $\mb{E}^{(1,1)}$ acts on the pair
$\mb{\metb}^{(1,1)}$,  $ \mb{\psi}_A^{(1,1 )}$ in Eq.~(\ref{EinCoup}) in the same way that
 $\mb{E}^{(0,1)}$
acts on $\mb{\metb}^{(0,1)}$, $ \mb{\psi}_A^{(0,1 )}$  in Eq.~(\ref{Enonr}).
The reason for this is that both operators $\mb{E}^{(0,1)}$ and $\mb{E}^{(1,1)}$ come from
the linearization, around the static background,  of the Einstein tensor operator acting
on non--radial perturbations.  Therefore, using again the linearity of $\mb{E}^{(1,1)}$,
we can define
\begin{equation}
\mb{L}_{\rm NR}\left[\,\cdot\,\right] \equiv \mb{E}^{(1,1)}\left[\,\cdot\, |\, 
\mb{0}\,\right]  = \mb{E}^{(0,1)}\left[\,\cdot\,\right]  \,,
\end{equation}
as the non-radial perturbation operator. Hence,
we can re-write equation
(\ref{EinCoup}) in the final form
\begin{equation}
\mb{L}_{\rm NR}\left[\,\mb{\metb}^{(1,1)}\,,\mb{\psi}_A^{(1,1)}\,\right]= \mb{S}\left[
\mb{\metb}^{(1,0)}\otimes\mb{\metb}^{(0,1)}\,, \mb{\psi}_A^{(1,0)}\otimes\mb{\psi}_A^{(0,1)}\,,
\mb{\metb}^{(1,0)}\otimes\mb{\psi}_A^{(0,1)}\,, \mb{\psi}_A^{(1,0)}\otimes\mb{\metb}^{(0,1)} \,\right]\,.
 \label{Einsep}
\end{equation}
The particular structure of these equations is very helpful
in order to develop a numerical code for studying the coupling of radial
and non-radial perturbations. We want to emphasize here that our goal is that of  solving
the  perturbative equations in the  time domain.
To this end it is then very useful to rely on well known initial value formulations for
the linear non-radial perturbations. Thus,
 assuming we have  working numerical
codes for the first order perturbations in $\lambda$ and
$\epsilon$,  in order to implement a code
for the coupling  $\lambda\epsilon$ variables we only need to modify the  code
for the $\epsilon$ variables by adding the source terms  to the right-hand side
of the evolution algorithm.
Then, at every time step in the evolution, and having fixed the
background, we have to: {\it i)} evolve the
equations to obtain the value of the first order
radial and non-radial perturbations; {\it ii)} to use their values
to evolve the coupling variables.

\section{Summary of the GSGM formalism} \label{sec:GSGM}
In this Section we briefly recall the GSGM formalism, introduced by Gerlach and Sengupta
\cite{Gerlach:1979rw,Gerlach:1980tx} and further developed by Gundlach and
Mart\'{\i}n--Garc\'{\i}a~\cite{Martin-Garcia:1998sk,Gundlach:1999bt,Martin-Garcia:2000ze}, to study
first order gauge-invariant perturbations of a general time-dependent
spherically symmetric stellar background. From the formal point of view of parameter
expansion of Section \ref{framework}, here we are dealing with standard 1-parameter linear
perturbations of the form
\begin{equation}
\met_{\alpha\beta} = \met^{(0)}_{\alpha\beta} + \epsilon \,
\met^{(1)}_{\alpha\beta}\,,\label{reorg}
\end{equation}
where $\mb{\met}^{(0)}$ is the  time-dependent
spherically symmetric  background and the $\epsilon$ perturbations are non-radial.

\subsection{The time dependent perfect fluid background}

The background manifold is the warped product $M^2\times S^2$, where
$S^2$ denotes the 2-sphere and $M^2$ a two-dimensional Lorentzian
manifold.  The metric can be written as the semidirect product of a
general Lorentzian metric on $M^2$, $g_{AB}$, and the unit curvature
metric on $S^2$, that we call $\gamma_{ab}$:
\begin{equation}
\met_{\alpha\beta} = \left(\begin{array}{cc}
g_{AB} & 0 \\
0 & r^2\gamma_{ab} \end{array} \right) \,.\label{met22}
\end{equation}
Hereafter, $x^A$ denotes the coordinates on $M^2$ and
$x^a$ the coordinates on $S^2$; $r=r(x^A)$ is a function on $M^2$ that
coincides with the invariantly defined radial (area) coordinate of
spherically-symmetric spacetimes.
A vertical bar is used to denote the
covariant derivative on $M^2$ and a semicolon to denote the one on
$S^2$, thus we have $g_{AB|C}=\gamma_{ab:c}=0\,.$  One can introduce the
completely antisymmetric covariant unit tensors on $M^2$ and on $S^2$,
$\epsilon_{AB}$ and $\epsilon_{ab}$ respectively, in such a way they
satisfy:
$\epsilon_{AB|C}=\epsilon_{ab:c}=0\,,$
$\epsilon_{AC}\epsilon^{BC}= -g_A^B\,,$ and
$\epsilon_{ac}\epsilon^{bc}= -\gamma_a^c\,.$

In this paper we consider a perfect-fluid description of the stellar
matter, thus the energy-momentum tensor is
\begin{equation}
t_{\alpha\beta}=(\rho+p)u_{\alpha}u_{\beta}+p\met_{\alpha\beta}\,,
\end{equation}
where $\rho$ and $p$ are the energy density and pressure, and
$u_{\alpha}$ is the fluid velocity.

In the spherically symmetry background $t_{\alpha\beta}$ has the same block
diagonal structure than the metric,
\begin{equation}
t_{\alpha\beta}= \textrm{diag} \left( \
t_{AB}\;, \,r^2 Q(x^C) \gamma_{ab} \right)\,, \label{tblock}
\end{equation}
and the fluid velocity takes the form
$u_{\alpha}=( u_A,0)$.
An orthonormal frame on the submanifold $M^2$ can be formed from
$u^A$ and the spacelike vector
\begin{equation}
n_A\equiv-\epsilon_{AB}u^B~~~\Rightarrow~~~n_A u^A=0\,.
\end{equation}
The metric $\met_{AB}$ and $\epsilon_{AB}$ can be written
in terms of these frame vectors as follows
\begin{eqnarray}
g_{AB} =  -u_Au_B+n_An_B\,,~~~~ \qquad
\epsilon_{AB} = n_Au_B-u_An_B\,. \label{ge22}
\end{eqnarray}
Then we have
\begin{eqnarray}
t_{AB}& = &\rho u_Au_B+pn_An_B\,,~~~~Q = p\,.
\end{eqnarray}
In any given coordinate system for $M^2$, $\{x^A\}\,,$ one can define the
following quantity:
\begin{equation}
v_A\equiv \frac{1}{r}r_{|A}\,.
\end{equation}
Then, any covariant derivative on the spacetime can be written in
terms of the covariant derivatives on $M^2$ and $S^2$, plus some terms
due to the warp factor $r^2$, which can be written in terms of
$v_A$.
Finally, the frame derivatives of a generic scalar function $f$ are defined by
\begin{equation}
\dot f=u^Af_{|A}\,,~~~f'=n^Af_{|A}\,,
\end{equation}
and we introduce the following background scalars:
\begin{equation}
\Omega=\ln\rho,~~~U=u^Av_A,~~~W=n^Av_A,~~~\mu=u^A_{~|A},~~~\nu=n^A_{~|A}\,.
\end{equation}

\subsection{Perturbations}

Linear perturbations of a spherically-symmetric background can
be decomposed in scalar, vector and tensor spherical harmonics. The
scalar spherical harmonics $Y^{lm}$ are eigenfunctions of the
covariant Laplacian on the sphere:
\begin{equation}
\gamma^{ab}Y^{lm}_{:ab}=-l(l+1)Y^{lm}\,.
\end{equation}
A basis of vector spherical harmonics (defined for $l\ge 1$) is
\begin{equation}
Y^{lm}_a\equiv Y^{lm}_{:a}\,,~~~S^{lm}_a\equiv \epsilon_a^bY^{lm}_b\,,
\end{equation}
where the $Y_a^{lm}$'s have polar parity (they transform as $(-1)^l$,
like the scalar harmonics, under parity transformations, and are also called
even-parity type) and the
$S_a^{lm}$'s have axial parity (they transform as $(-1)^{l+1}$ under
parity transformations, and are also called odd-parity type). A basis of tensor
spherical harmonics (defined for $l\ge 2$) is
\begin{equation}
Y_{ab}^{lm}\equiv Y^{lm}\gamma_{ab}\,,~~~
Z^{lm}_{ab}\equiv Y^{lm}_{:ab}+\frac{l(l+1)}{2}Y^{lm}\gamma_{ab}\,,
~~~S^{lm}_{ab}\equiv S^{lm}_{a:b}+S^{lm}_{b:a}\,,
\end{equation}
where the $Y_{ab}^{lm},\,Z_{ab}^{lm}$ have polar parity and the $S^{lm}_{ab}$
have axial parity.
In this paper we will only consider perturbations with polar parity.

The perturbations of the covariant metric and energy-momentum tensors can be expanded in
this basis as
\begin{eqnarray}
\delta g_{\alpha\beta} & = & \left(\begin{array}{cc}
  h_{AB}^{lm}\, Y^{lm} &  h_A^{lm} \, Y^{lm}_a \\
\\
  h_A^{lm}  \, Y^{lm}_a    & \  r^2(K^{lm} \, \gamma_{ab} \, Y^{lm}+
  G^{lm} \, Y_{:ab}^{lm})\\
\end{array}\right)\,, \label{newlab} \\
\delta t_{\alpha\beta}&=&\left(\begin{array}{cc}
  \delta t_{AB}^{lm} \, Y^{lm} &  \delta t_A^{lm} \, Y^{lm}_a \\
\\
  \delta t_A^{lm} \, Y^{lm}_a & \ r^2 \, \delta t^{3\,lm} \, \gamma_{ab} \, Y^{lm}+
\delta t^{2\,lm} \, Y_{:ab}^{lm} \\
\end{array}\right)\,. \label{talbe}
\end{eqnarray}
Let $X$ be an arbitrary tensor field on the background spacetime and
$\delta X$ its linear perturbation.  It is well-known that under a a first order gauge
transformation,  generated by a vector field $\xi$ living on the background,
the perturbation of $X$ transforms as
\begin{equation}
\delta X\rightarrow\delta X+{\cal L}_{\xi}X\,. \label{GIFO}
\end{equation}
Then, the perturbation $\delta X$ is gauge-invariant if and only if
the Lie derivative of the corresponding background quantity $X$ with
respect to an arbitrary vector field $\xi$ vanishes:
${\cal L}_{\xi}X=0$~\cite{1974RSPSA.341...49S}.  Using this well-known result,
it is possible to show that a complete set of gauge-invariant variables,
combinations of the
perturbations $h_{AB},\,h_A,\,K,\,G$, $ \delta t_{AB},\,\delta
t_A,\,\delta t^2,\,\delta t^3\,,$ is given by the following quantities:
\begin{eqnarray}
k_{AB}&=&h_{AB}-(p_{A|B}+p_{B|A})\,, \label{pkab} \\
k&=&K-2v^Ap_A\,, \label{pk} \\
T_{AB}&=&\delta t_{AB}-t_{AB|C}\, p^C-t_{AC} \,  p^C_{|B}-t_{BC} \, p^C_{|A}\,, 
\label{pTAB}\\
T^3&=&\delta t^3-p^C(Q_{|C}+2Qv_C)+\frac{l(l+1)}{2} \, Q \, G\,, \\
T_A&=&\delta t_A-t_{AC} \, p^C-\frac{r^2}{2} \, Q \, G_{|A}\,, \\
T^2&=&\delta t^2-r^2\, Q\, G \,, \label{pt2}
\end{eqnarray}
where $T_A$ is defined for $l\ge 1\,,$  $T^2$ is defined for $l\ge 2$, and
\begin{equation}
p_A = h_A - \frac{1}{2}r^2G_{|A}\,. \label{ppa}
\end{equation}
Therefore, any linear perturbation of the spherically-symmetric background
(\ref{met22}) can be written as a linear combination of these gauge-invariant 
quantities.
Is is important to stress the fact that although we are considering a perfect fluid
energy-momentum tensor, the gauge invariant quantities introduced above are
defined for any energy-momentum tensor.   The equation of state for a perfect
fluid has the form $p=p(\rho,s)$,  $s$ being the specific entropy.   The
corresponding sound speed, $c_s$, can then be defined through the
 thermodynamical derivative
\begin{equation}
c_s^2=\left(\frac{\partial p}{\partial\rho}\right)_s\,.
\end{equation}
Throughout this paper we will consider the particular case of a barotropic fluid
(i.e. constant specific entropy), hence we will have $p=p(\rho)$.

The polar fluid perturbations are described as follows.  The
perturbations of the fluid velocity components can be written as
\begin{equation}
\delta u_{\alpha}=\left(\left[\tilde\gamma^{lm} n_A+
\frac{1}{2}h_{AB}^{lm}u^B\right]Y^{lm}\,,
\tilde\alpha^{lm} Y^{lm}_{:a}\right)\,, \label{deltau}
\end{equation}
where $\tilde\alpha$ is defined for $l\ge 1\,.$  The energy
density and pressure perturbations can be cast in the following
form (using the barotropic equation of state)
\begin{equation}
\delta\rho=\tilde\omega\rho Y^{lm} \,,~~~~\delta
p=c_s^2\delta\rho\,.  \label{pr}
\end{equation}
In terms of these quantities it is possible to define a
gauge-invariant set of fluid perturbations:
\begin{eqnarray}
\alpha&=&\tilde\alpha-p^Bu_B\,, \label{algi} \\
\gamma&=&\tilde\gamma-n^A\left[p^Bu_{A|B}
+\frac{1}{2}u^B(p_{B|A}-p_{A|B})\right]\,, \label{gamgi}  \\
\omega&=&\tilde\omega-p^A\Omega_{|A}\,.   \label{omgi}
\end{eqnarray}
The tensor $k_{AB}$ can be decomposed in the frame $\{u^A,n^A\}$:
\begin{equation}
k_{AB}=\eta(-u_Au_B+n_An_B)+\phi(u_Au_B+n_An_B)+\psi(u_An_B+n_Au_B)\,, \label{KABframe}
\end{equation}
where $\eta$, $\phi$ and $\psi$ are scalars.  It is useful to consider
the following new scalar variable
\begin{equation}
\chi=\phi-k+\eta\,, \label{chidef}
\end{equation}
in the place of $\phi$.  Then, combining Einstein equations with the 
energy-momentum equations we can obtain the following set of equations:
for $l\ge 2$,  
\begin{equation}
\eta=0\,, \label{eta}
\end{equation}
for $l\ge 1$,
\begin{eqnarray}
-\ddot\chi+\chi''+2(\mu-U)\psi'&=&S_{\chi}\,, \label{chitt} \\
-\ddot k+c_s^2k''-2c_s^2U\psi'&=&S_k\,, \label{ktt} \\
-\dot\psi&=&S_{\psi}\,, \label{psit} \\
16\pi(\rho+p)\alpha&=&\psi'+C_{\alpha}\,, \label{psip}\\
-\dot\alpha&=&S_{\alpha}\,, \label{alphat}  \\
-\dot\omega-\left(1+\frac{p}{\rho}\right)\gamma'&=&\bar S_{\omega}\,,
\label{omegat} \\
\left(1+\frac{p}{\rho}\right)\dot\gamma+c_s^2\omega'&=&\bar S_{\gamma}  
\label{gammat} \,.
\end{eqnarray}
And finally, for $l\ge 0$,
\begin{eqnarray}
8\pi(\rho+p)\gamma&=&(\dot k)'+C_{\gamma}\,, \label{ktp} \\
8\pi\rho\omega&=&-k''+2U\psi'+C_{\omega}\,, \label{kpp} \, .
\end{eqnarray}
where the expressions of
$S_{\chi},\,S_{\psi},\,C_{\alpha},\,S_{\alpha},\, \bar
S_{\omega},\,\bar S_{\gamma},\,C_{\gamma},\,C_{\omega}$ can be found
in~\cite{Gundlach:1999bt}.

\section{Radial and Non-radial Perturbations of Static Relativistic Stars}
\label{sec:rad_MTW}

In what follows we  summarize the first order perturbative analysis of
the oscillations of a perfect-fluid static star.  As anticipated in Section \ref{framework},
we consider separately the radial pulsations, parametrized by $\lambda$,  and the non-radial
oscillations, parametrized by $\epsilon$. Therefore, consistently with the notation of
Section~\ref{framework}, we will now explicitly use the indices $(i,j)$; however,  to
simplify the notation, from now on we will use a bar to denote
quantities associated with the static spacetime, the background of our
2-parameter perturbative formalism. Thus we have
$\bar{\met}_{\alpha\beta}\equiv \met^{(0,0)}_{\alpha\beta}$, and in the same way
$\bar{\rho}$, $\bar{p}$, and $\bar{u}_\alpha$.

The equilibrium configuration is described by the static spherically-symmetric
metric:
\begin{equation}
\bar{\met}_{\alpha\beta}dx^\alpha dx^\beta =
-e^{2\Phi(r)}dt^2+e^{2\Lambda(r)}dr^2+
r^2(d\theta^2+\sin^2\theta d\phi^2)\,,
\end{equation}
with a perfect-fluid energy-momentum tensor
\begin{equation}
\bar{t}_{\alpha\beta} = (\bar{\rho}+\bar{p})\, \bar{u}_\alpha
\bar{u}_\beta + \bar{p}\, \bar{\met}_{\alpha\beta}\,,
\end{equation}
where $\bar{u}_\alpha=\left(-e^\Phi,0,0,0\right)$.  The mass function
is introduced by means of the equality $e^{-2\Lambda(r)}=1-2M(r)/r\,.$ Then,
the TOV equations are:
\beq
& & \Phi_{,r}=\frac{M+4\pi \bar{p} \, r^3}{r\, (r-2M)} =
-\frac{\bar{p}_{,r}}{\bar{\rho}+\bar{p}}\,, \\
& & M_{,r}=4\pi\bar{\rho} \, r^2 \,.
\eeq
Specifying the equation of state of the stellar matter one obtains a 1-parameter family
of solutions of these equations, depending on the central density. For the barotropic
equation of state we use, $\bar{p}=\bar{p}(\bar{\rho})$, the background sound speed is
$\bar{c}_s^2= d\bar{p}/ d\bar{\rho}\,.$

\subsection{Non-radial perturbations in the GSGM formalism} \label{Non-rad}

The equations for the first order non-radial perturbations have been
known for a long time~\cite{Thorne:1967th,Detweiler:1985dl,Chandrasekhar:1991fi}.
In~\cite{Nagar:2004ns} they have been presented in the framework of the
GSGM formalism.  We do not write their explicit expressions here,
because they can be
obtained as a particular case of the equations that we will write in
Section~\ref{secVc} for the coupling perturbative terms of order $(1,1)$, like
$\mb{\metb^{(1,1)}}$,  considering the homogeneous part of those equations, i.e.\ simply
neglecting the source terms.
One can also obtain the same equations directly  from the general GSGM
equations (\ref{eta}-\ref{kpp}),  by considering the special case of a static background,
represented by  the following quantities,
\begin{eqnarray}
\bar u^A = (e^{-\Phi},0)\,,~~~~ \bar n^A = (0,e^{- \, \Lambda})\,,  \label{unst}  \\
\bar\mu = \bar U =0\,,~~~~ \bar\nu = \Phi'\,,~~~~ \bar W =
\frac{e^{-\Lambda}}{r}\,, \label{mu_W_st}
\end{eqnarray}
and where the frame derivatives of a scalar function $f$  take the special form
$f' = e^{-\Lambda}f_{,r}$  and $\dot{f} = e^{-\Phi}f_{,t}$.
Boundary conditions have to be imposed at infinity, at the stellar origin
and on the stellar surface. They can be found in the references cited above.

\subsection{Radial perturbations in the GSGM formalism}

Radial perturbations can be seen as a particular subcase of non-radial
ones, namely those corresponding to the harmonic $l=0$.
The fact that the GSGM quantities are not gauge-invariant in the $l=0$
case does not represent a problem for our study since, when we will consider the gauge
invariance of our $(1,1)$ variables in Section~\ref{GI11},   we will assume that the gauge
of  the perturbative order $(1,0)$ has been somehow fixed.  In practice, we
fix the radial gauge as in~\cite{Gundlach:1999bt},
\begin{equation}
\psi^{(1,0)}=0\,,
\qquad k^{(1,0)}=0\,, \label{fixgauge}
\end{equation}
which considerably simplifies
 the equations.  In this gauge, the metric perturbations
have a diagonal form as in the background:
\begin{equation}
\met_{\alpha \beta}^{(1,0)}=
\mbox{diag}\left(h_{AB}^{(1,0)}, 0 \right) \,, \label{eq:rad_metr}
\end{equation}
\begin{equation}
h_{AB}^{(1,0)} = \eta^{(1,0)} \bar{\met}_{AB} + \phi^{(1,0)}
(\bar{u}_A\bar{u}_B+\bar{n}_A\bar{n}_B) \,.
\end{equation}
Then, the components of $h_{AB}^{(1,0)}$ are given by
\begin{equation}
h_{AB}^{(1,0)}=
\mbox{diag} \left( e^{2\Phi}\left( \chi^{(1,0)} - 2\eta^{(1,0)} \right)\,,
e^{2\Lambda}\,\chi^{(1,0)} \right) \,,
\end{equation}
where $\chi^{(1,0)}= \eta^{(1,0)} + \phi^{(1,0)}$.

For the fluid velocity, from Eq.~(\ref{deltau}), we have
\begin{equation}
\delta u_A^{(1,0)}= \left(  \left( \frac{\chi^{(1,0)}}{2} - \eta^{(1,0)}
\right) e^{\Phi}\,, \, e^{\Lambda} \,\gamma^{(1,0)}
\right)\,,~~~~ \d u_a^{(1,0)} = 0\,.
\end{equation}
The other fluid perturbations are given by:
\begin{equation}
\delta \rho^{(1,0)}=\omega^{(1,0)}
\bar\rho\,,  \qquad \delta p^{(1,0)} = \bar{c}_s^2  \delta
\rho^{(1,0)} \,.
\end{equation}

\subsection{Equations for the radial perturbations}

The equations for the quantities $\chi^{(1,0)},
\eta^{(1,0)},\omega^{(1,0)},\gamma^{(1,0)}$ describing the  radial
 perturbations can be found in~\cite{Gundlach:1999bt}.  For our
purposes it is more convenient to use a different set of variables.
First, instead of using $\omega^{(1,0)}$ we use the enthalpy
perturbation $H^{(1,0)}$, which significantly simplifies  the equations.
The second
change consists in replacing the metric perturbation $\chi^{(1,0)}$
with the quantity $S^{(1,0)}$, in order to use a set of variables
consistent with the one we will use for the non-radial perturbations
in Section~\ref{secVc}.  The definitions of $H^{(1,0)}$ and
$S^{(1,0)}$ are:
\begin{equation}
H^{(1,0)} \equiv \frac{\d p^{(1,0)}}{\bar{\rho}+\bar{p}} =
\frac{\bar{c}_s^2 \bar{\rho}}{\bar{\rho} +\bar{p}}\,\om^{(1,0)}\,,  \qquad
S^{(1,0)} \equiv \frac{\chi^{(1,0)}}{r} \,.
\end{equation}
Furthermore, using equation (34) in~\cite{Martin-Garcia:1998sk} we end up with
a set of evolution equations for $S^{(1,0)},H^{(1,0)}$ and
$\gamma^{(1,0)}$ which does not contain the quantity $\eta^{(1,0)}$,
\begin{eqnarray}
-\dot{H}^{(1,0)}  & = &  \bar{c}_s^2\,\ga^{(1,0)'} + \bar{c}_s^2 \,
\left[\left(1-\frac{1}{\bar{c}_s^2}\right) \bar{\nu} + 2\bar{W} -
\frac{4\pi}{\bar{W}}\left(\bar\rho + \bar p \right)  \right]\, \ga^{(1,0)}\,,
\label{eq:H10_ev}  \\
\dot{\ga}^{(1,0)} & = &  - H^{(1,0)'} - \frac{4\pi}{\bar{W}}\,
\left(\bar\rho + \bar p\right)\, H^{(1,0)} -
\left(\bar\nu+\frac{\bar W}{2} \right) r\,S^{(1,0)}\,, \label{eq:S10_ev} \\
\dot{S}^{(1,0)}  & = &  -8 \pi \left(\bar\rho + \bar p \right)\,
\frac{1}{r \bar W}\,\ga^{(1,0)}\,, \label{chi_t}
\end{eqnarray}
and with the Hamiltonian constraint
\begin{eqnarray}
\bar W\,r\,S^{(1,0)'} & = &  \left( 8 \pi \bar\rho r -\frac{1}{r} - \bar W r'
\right)\, S^{(1,0)}+8 \pi \, \frac{\bar\rho + \bar p}{\bar c_s^2} \, 
 H^{(1,0)}   \,.\label{eq:S10_cn}
\end{eqnarray}
The quantity $\eta^{(1,0)}$ can be found, in terms of
$(S^{(1,0)},H^{(1,0)},\gamma^{(1,0)})$, from the following equation:
\begin{equation}
\bar W \eta^{(1,0)'} = 4\pi (\bar\rho+\bar
p)\left[r\,S^{(1,0)}+\left(1+
\frac{1}{\bar{c}^2_s}\right)H^{(1,0)}  \right] \,.
\label{eq:eta_cn}
\end{equation}
The system of equations (\ref{eq:H10_ev}-\ref{eq:S10_cn}) is
equivalent to the one used by Ruoff~\cite{Ruoff:2000nj}, cf.\ also
\cite{Misner:1973cw}.

\subsection{Boundary Conditions for radial perturbations}

Boundary conditions must be fixed at the center and at the stellar surface $r=R$.
The latter is given by  the vanishing of the Lagrangian
pressure perturbation, $\Delta p=0$.  Following~\cite{Misner:1973cw},
we can express $\Delta p^{(1,0)}$ in terms of the {\em radial renormalized
displacement function} $\zeta$:
\be
r^2 \D p^{(1,0)}= -\left(\bar\rho + \bar p \right)  \bar{c}_s^2 e^{-\Phi}
\frac{\partial \zeta}{\partial r} \,.
\ee
Using the relation between  $\zeta$ and $\gamma^{(1,0)}$
\begin{equation}
\zeta_{,t} = r^2 e^{-\Lambda} \gamma^{(1,0)} \,
\end{equation}
we arrive at the following boundary condition on the surface:
\begin{equation}
\left. (\bar\rho + \bar p)\,\bar{c}^2_s\, e^{-\Phi}\left( r^2 e^{-\Lambda} 
\gamma^{(1,0)}\right)_{,r} \right|_{r=R}= 0\,.
\end{equation}
The  behaviour  of  $S^{(1,0)}$ and $H^{(1,0)}$ on the surface can be found from
the general evolution equations~(\ref{chi_t}) and~(\ref{eq:H10_ev}).

The analysis of the regularity conditions at the origin ($r=0$)
leads to the following expressions:
\be
(S^{(1,0)},\eta^{(1,0)},H^{(1,0)},\gamma^{(1,0)})~~\longrightarrow~~
(S^{(1,0)}_o(t)\,r+ O(r^3)\,, \eta^{(1,0)}_o(t)+O(r^2)\,,
H^{(1,0)}_o(t)+O(r^2)\,, \gamma^{(1,0)}_o(t)\,r+ O(r^3)\,)\,.
\label{bc_rad}
\ee

\section{Coupling of radial and non-radial perturbations of relativistic stars}
\label{coupsec}
 
At this point we have already specified the equations that
determine the static background metric $\mb{\metb}^{(0,0)}$, the radial
perturbations of order $\lambda$, and the non-radial perturbations of order $\epsilon$.
The next crucial step is to find the equations for
the perturbative coupling terms of order $\lambda\epsilon$, like the metric perturbations $\mb{\metb}^{(1,1)}$.
To this end we can further expand the first order 1-parameter GSGM formalism of Section~\ref{sec:GSGM}
as follows. The time dependent spherically symmetric variables can be split into a static and a radially
oscillating parts, e.g.\ for the metric
\begin{equation}
\met^{(0)}_{\alpha\beta}  =  \met^{(0,0)}_{\alpha\beta} + \lambda \,
\met^{(1,0)}_{\alpha\beta}\,. \label{bgsep}
\end{equation}
The non-radial first order perturbations on this time dependent spacetime can be split into a part
which is the non-radial perturbation of the static background and a further term that describes the
coupling. For the metric we then have
\begin{equation}
\met^{(1)}_{\alpha\beta}  = \met^{(0,1)}_{\alpha\beta} + \lambda \,
\met^{(1,1)}_{\alpha\beta}\,. \label{fosep}
\end{equation}
Inserting (\ref{bgsep}) and (\ref{fosep}) into Eq.~(\ref{reorg}) we re-obtain the 2-parameter
expansion (\ref{initialmetric}), and similarly for the fluid variables.

This approach, based on  splitting the GSGM variables and their equations, is very convenient
because  it saves us from the need of computing directly a
second order expansion of Einstein's equations.
In addition, following  this approach we shall  show in Section \ref{GI11} how to build
gauge-invariant quantities
associated with the perturbations $\metb^{(1,1)}$ and the fluid variables of order $(1,1)$.

\subsection{GSGM formalism on a radially oscillating star}

In order to implement   the GSGM formalism  on a  radially oscillating star, which is itself
treated perturbatively, we split  the quantities $u^A,\, n^A,\, U,\,
W,\, \mu,\, \nu$ in a static background part and a radial perturbation, as illustrated above.
The frame vector fields are given by
\begin{eqnarray}
u^A & = & \left( \left[ 1- \la  \left( \eta^{(1,0)} - \frac{\chi^{(1,0)}}{2}
 \right)  \right]  e^{-\Phi} ,  \la \
e^{-\Lambda}   \ga^{(1,0)}  \right) \,,  \label{urdbg} \\
n^A & = & \left( \la  e^{-\Phi}  \ga^{(1,0)} ,  \left(1-
\la  \frac{\chi^{(1,0)}}{2} \,  \right) e^{-\Lambda}\right) \,.  \label{nrdbg}
\end{eqnarray}
The frame derivatives of a scalar perturbation,
$f^{(1)}=f^{(0,1)}+\lambda f^{(1,1)}$, on the radially oscillating star are:
\begin{eqnarray}
\dot{f}^{(1)} & = & u^A f^{(1)}_{,A} =  e^{-\Phi}
f^{(0,1)}_{,t} + \la \left\{e^{-\Phi} f^{(1,1)}_{,t} +
 e^{-\Lambda}  \ga^{(1,0)} f^{(0,1)}_{,r} -
e^{-\Phi}  \left( \eta^{(1,0)} -  \frac{\chi^{(1,0)}}{2}
\right) f^{(0,1)}_{,t} \, \right\}  \,, \\
f^{(1)'} & = & n^A f^{(1)}_{,A}  = e^{-\Lambda}
f^{(0,1)}_{,r} + \la \left\{e^{-\Lambda} f^{(1,1)}_{,r} +
e^{-\Phi} \ga^{(1,0)} f^{(0,1)}_{,t} - \frac{\chi^{(1,0)}}{2}
 \, e^{-\Lambda} f^{(0,1)}_{,r} \right\} \,.
\end{eqnarray}
The remaining  quantities describing the spherical star are
\beq
U & = & u^Av_A = \lambda \frac{e^{-\Lambda}}{r}   \ga_{,t}^{(1,0)}\,,   \label{Ubk} \\
W & = & n^Av_A = \left( \frac{1}{r}-\lambda   \frac{\chi^{(1,0)}}{2 \, r}
\right) \ e^{-\Lambda}\,,  \label{Wbk} \\
\mu & = &  u^A_{~|A}  = \lambda \, \left( \ga^{(1,0)} \, e^{-\Lambda}
\right)_{,\, r}  \label{mubk} \,,\\
\nu & = & n^A_{~|A} = \Phi_{,r}  e^{-\Lambda} + \lambda\,\left\{e^{-\Phi}
\ga_{,t}^{(1,0)}
+ e^{-\Lambda}\left[\left( \eta_{,r}^{(1,0)} -\frac{1}{2}\chi_{,r}^{(1,0)}\right)
-\frac{1}{2} \Phi_{,r} \chi^{(1,0)} \right]\right\}\, , \label{nubk}
\eeq
where in (\ref{mubk}) we have used the (\ref{chi_t}).

\subsection{Gauge invariance of the $\lambda\epsilon$ coupling perturbations} \label{GI11}

In this Section we are going to construct a set of gauge-invariant quantities at
the $(1,1)$ perturbative order.  As we shall see,  we can find
them with the help of the GSGM formalism.  The main idea behind our
construction is to build the $(1,1)$ gauge-invariant variables
starting from the gauge-invariant quantities of the GSGM formalism,
considering these latter as perturbations of a
radially pulsating spacetime which is itself described as a perturbation of the static
background, i.e. $\met^{(0)}_{\alpha\beta} =
\met^{(0,0)}_{\alpha\beta}+\lambda\met^{(1,0)}_{\alpha\beta}$.  More
precisely, let ${\cal G}^{(1)}$ be any of the gauge-invariant
quantities  in equations (\ref{pkab}-\ref{pt2}) and  (\ref{algi}-\ref{omgi}), but constructed
as a metric and energy-momentum perturbation of the pulsating
star.  Then, we can expand ${\cal G}^{(1)}$ in $\lambda$ to
get
\begin{equation}
{\cal G}^{(1)} = {\cal G}^{(0,1)} +\lambda{\cal G}^{(1,1)}\,. \label{gp1p}
\end{equation}
It is important to remark here that the $(1,1)$ superscript refers not
only to quantities constructed from the $\metb^{(1,1)}$ perturbations,
but in general to any perturbative quantity of order
$\lambda\epsilon$.  In general ${\cal G}^{(0,1)}$ and ${\cal
G}^{(1,1)}$ can be expressed as:
\begin{eqnarray}
{\cal G}^{(0,1)} & = & {\cal H}^{(0,1)} \,, \label{g01}\\
{\cal G}^{(1,1)} & = & {\cal H}^{(1,1)} + \sum_\sigma{\cal I}^{(1,0)}_\sigma
{\cal J}^{(0,1)}_\sigma\,,\label{g11}
\end{eqnarray}
where the objects ${\cal H}^{(0,1)}$ and  ${\cal J}^{(0,1)}_\sigma$ are linear in the ${(0,1)}$
perturbations, while  ${\cal I}^{(1,0)}_\sigma$   and ${\cal H}^{(1,1)}$ are respectively  linear in the
${(1,0)}$ and the ${(1,1)}$ variables.   It is clear that ${\cal H}^{(0,1)}$ is
nothing  but any of the  gauge-invariant quantities in
(\ref{pkab}-\ref{pt2}) and (\ref{algi}-\ref{omgi}) for  the special case of a static background,
as it must be for the $(0,1)$
perturbations.  The quantity ${\cal H}^{(1,1)}$ is constructed from  $(1,1)$ quantities in  the same
way but, as we are going to see, it is not a gauge-invariant object at
the order $(1,1)$: we have to add some extra terms of the form
given in (\ref{g11}) and which come from the ansatz (\ref{gp1p}).  In
what follows we show that this gives the desired gauge-invariant
$(1,1)$ quantities.

Gauge transformations and gauge invariance in 2-parameter
perturbation theory have been studied
in~\cite{Bruni:2002sm,Sopuerta:2003rg}.   The gauge
transformation of a  first order perturbation of order $(0,1)$
of a generic tensorial quantity ${\cal T}$ is given, as in~(\ref{GIFO}),  by
\begin{eqnarray}
\tilde{\cal T}^{(0,1)}& =& {\cal T}^{(0,1)}
+ \pounds_{\xi_{(0,1)}} {\cal T}^{(0,0)}\,, \label{gafo}
\end{eqnarray}
whereas a  second order perturbation of ${\cal T}$, in particular of order $(1,1)$, transforms
according to
\begin{eqnarray}
\tilde{\cal T}^{(1,1)} = {\cal T}^{(1,1)} +
\pounds_{\xi_{(0,1)}}\,{\cal T}^{(1,0)}
+\pounds_{\xi_{(1,0)}}\,{\cal T}^{(0,1)} +
\left(\pounds_{\xi_{(1,1)}} +
\left\{\pounds_{\xi_{(1,0)}}\,,\,\pounds_{\xi_{(0,1)}}
\right\} \right) {\cal T}^{(0,0)} \,,
\end{eqnarray} where
$\{\,,\}$ stands for the anti-commutator
$\{a,b\}= a\, b + b\, a$.

In the present case, we have chosen to fix the gauge for the radial
perturbations $\metb^{(1,0)}$, see equation~(\ref{fixgauge}).  This
simplifies  the previous transformation rule  to
\begin{equation}
\tilde{\cal T}^{(1,1)} = {\cal T}^{(1,1)} +
\pounds_{\xi_{(0,1)}}\,{\cal T}^{(1,0)}
+\pounds_{\xi_{(1,1)}}\,{\cal T}^{(0,0)} \,. \label{ga_tran}
\end{equation}
We have  assumed that ${\cal G}^{(0,1)}$is a  gauge-invariant quantity at
order $(0,1)$, therefore
\begin{equation}
\tilde{\cal G}^{(0,1)}-{\cal G}^{(0,1)} = \tilde{\cal H}^{(0,1)}-{\cal H}^{(0,1)}
=0\,. \label{agi01}
\end{equation}
{}From~(\ref{g11}) and the fact that we have fixed the gauge for radial
perturbations, we can write
\begin{eqnarray}
\tilde{\cal G}^{(1,1)}-{\cal G}^{(1,1)}
 & = & \tilde{\cal H}^{(1,1)} -{\cal H}^{(1,1)}
+ \sum_\sigma
{\cal I}^{(1,0)}_\sigma \, \left(\tilde{\cal J}^{(0,1)}_\sigma-{\cal J}^{(0,1)}_\sigma
\right)\,. \label{agi11}
\end{eqnarray}
Furthermore, we note that every ${\cal H}^{(1,1)}$ and ${\cal J}^{(0,1)}_{\sigma}$ can
be expressed as follows:
\begin{equation}
{\cal H}^{(1,1)} = {\cal A}\left[\metb^{(1,1)} \right]\,,~~~~
{\cal J}^{(0,1)}_\sigma = {\cal B}_\sigma\left[\metb^{(0,1)} \right]\,, \label{hjlin}
\end{equation}
where ${\cal A}$ and ${\cal B}_\sigma$ are linear operators involving
differentiation with respect to the coordinates of $M^2$ and
integration on $S^2$.  These operators act on spacetime objects and
return objects with indices on $M^2$.  From now on, for
the sake of simplicity, we only consider metric perturbations.
The corresponding procedure for energy-momentum tensor perturbations
follows the same lines and is given in Appendix \ref{AppGI}.  Using
the gauge transformations (\ref{gafo}) and (\ref{ga_tran}), the
transformation rules for ${\cal H}^{(1,1)}$ and ${\cal
J}^{(0,1)}_\sigma$ are given by
\begin{eqnarray}
\tilde{\cal H}^{(1,1)} &=&{\cal H}^{(1,1)}
 + {\cal A}\left[\pounds_{\xi_{(0,1)}}\metb^{(1,0)} +
\pounds_{\xi_{(1,1)}}\metb^{(0,0)} \right]  \,, \label{ha11}\\
\tilde{\cal J}^{(0,1)}_\sigma&=&{\cal J}^{(0,1)}_\sigma
 + {\cal B}_\sigma\left[ \pounds_{\xi_{(0,1)}}\metb^{(0,0)}
\right]\,.
\end{eqnarray}
Moreover, we know~\cite{Gerlach:1979rw,Gerlach:1980tx} that the
quantities in~(\ref{pkab}-\ref{pt2}) are gauge-invariant as
first order perturbations, hence ${\cal
A}\left[\pounds_{\xi}\metb^{(0,0)} \right]$ must vanish for any vector
field $\xi$.  Therefore, (\ref{ha11}) becomes
\begin{equation}
\tilde{\cal H}^{(1,1)}={\cal H}^{(1,1)}
 + {\cal A}\left[\pounds_{\xi_{(0,1)}}\metb^{(1,0)}\right] \,, \label{ha11_B}
\end{equation}
and the gauge transformation (\ref{agi11}) reduces to
\be
\tilde{\cal G}^{(1,1)}={\cal G}^{(1,1)}
 + {\cal A}\left[\pounds_{\xi_{(0,1)}}\metb^{(1,0)} \right] +
\sum_\sigma
{\cal I}^{(1,0)}_\sigma {\cal B}_\sigma \, \left[ \pounds_{\xi_{(0,1)}}\metb^{(0,0)}
\right] \,. \label{agi11_red}
\ee
Using this expression we can show that the variables ${\cal
G}^{(1,1)}$ are gauge-invariant.  For the sake of brevity we only
give here the proof for the metric perturbation
$k^{(1,1)}_{AB}$~(\ref{pkab}), the analysis for the other metric and
fluid perturbations in~(\ref{pkab}-\ref{pt2}) and (\ref{algi}-\ref{omgi})
is given in Appendix~\ref{AppGI}.  To proceed with the proof we expand the
generator of the gauge transformations associated with the non-radial
perturbations in tensor harmonics,
\begin{equation}
\xi_{(0,1)\alpha}=(\hat\xi_A \, Y\,,\, r^2 \xi \, Y_a )\,.
\label{xi01}
\end{equation}
Since the metric perturbations $\met^{(1,0)}$ do not depend on the coordinates of
$S^2$, and taking into account the gauge choice (\ref{fixgauge}), we have
\begin{equation}
\pounds_{\xi_{(0,1)}}\met_{AB}^{(1,0)}=\;
\hat{\!\!\pounds}_{\hat\xi}h_{AB}^{(1,0)}\,,~~~~
\quad \pounds_{\xi_{(0,1)}}\met_{Aa}^{(1,0)}= h_{A C}^{(1,0)}
\,\hat{\xi}^{C} \,Y_{:a} \,,~~~~ \quad
\pounds_{\xi_{(0,1)}}\met^{(1,0)}_{ab}=0\,,\label{Lieg10}
\end{equation}
where ${\!\!\pounds}_{\hat\xi}$ is the Lie derivative  acting
on $M^2$.  Now, we apply the ansatz described in (\ref{g11}) to
$k_{AB}$ and we get the following expressions
\begin{eqnarray}
{\cal H}^{(1,1)}_{AB} = h^{(1,1)}_{AB} -
\left(p^{(1,1)}_{A\mid B} + p^{(1,1)}_{B \mid A}\right)\,,~~~~
{\cal I}^{(1,0)C}_{AB} = 2\, \Gamma^{(1,0)C}_{AB}\,,~~~~
{\cal J}^{(0,1)}_C = p^{(0,1)}_C\, ,\label{HIJ}
\end{eqnarray}
where $p_{A}$ is defined in (\ref{ppa}) and $\Gamma^{(1,0)C}_{AB}$ are
the radial perturbations of the Christoffel symbols.  From (\ref{ha11_B})
and the analysis of ~\cite{Gerlach:1979rw,Gerlach:1980tx}, we have
\begin{equation}
\tilde{\cal H}^{(1,1)}_{AB}={\cal H}^{(1,1)}_{AB}
 +  \, \hat{\!\!\pounds}_{\hat\xi}\left[h^{(1,0)}_{AB}-
\left(p^{(1,0)}_{A\mid B}+p^{(1,0)}_{B\mid A}\right) \right] \,, ~~~~
\tilde{\cal J}^{(0,1)}_C={\cal J}^{(0,1)}_C
  +  \tilde p^{(0,1)}_C- p^{(0,1)}_C
 ={\cal J}^{(0,1)}_C+ \hat{\xi}_C\,, \label{HAB11}
\end{equation}
where the explicit expression of the Lie derivatives is
\begin{eqnarray}
\hat{\!\!\pounds}_{\hat\xi}h^{(1,0)}_{AB}  =  \hat{\xi}^C
h^{(1,0)}_{AB\mid C} + h^{(1,0)}_{CB}\hat{\xi}^C_{\mid A}
+h^{(1,0)}_{CA} \hat{\xi}^C_{\mid B}\,, ~~~~
\hat{\!\!\pounds}_{\hat\xi}p^{(1,0)}_A  =  h^{(1,0)}_{AC}
\hat{\xi}^C \,. \label{LiehAB}
\end{eqnarray}
Finally, introducing all the expressions (\ref{HIJ}-\ref{LiehAB}) into the gauge
transformation law (\ref{agi11_red}), we get the gauge invariance
of $k_{AB}$ at $(1,1)$ order:
\begin{equation}
\tilde{k}^{(1,1)}_{AB} = k^{(1,1)}_{AB} \,.
\end{equation}

\subsection{Equations for the $\lambda\epsilon$ coupling  perturbations}\label{secVc}

The explicit form of the equations that govern the behaviour of the coupling terms is
obtained by introducing in equations (\ref{chitt}-\ref{kpp}) the
following expressions: {\it i)} for the background quantities we will
use the expressions of the GSGM quantities describing the radially oscillating
spacetime (the static background plus radial perturbations), given by
equations~(\ref{urdbg}-\ref{nubk});  {\it ii)}  for the perturbative quantities
we use the corrections to the radially oscillating star, that is, the
 quantities that come from  perturbative terms like
$\met^{(1)}_{\alpha\beta} = \met_{\alpha\beta}^{(0,1)} +
\lambda \met_{\alpha\beta}^{(1,1)}\,.$
Once we have introduced all these quantities, expanded the equations and
extracted the $\lambda\epsilon$ part, we get a set of equations that
can be expressed as a linear non-radial operator $\mb{L}_{\rm NR}$ acting on the $(1,1)$ variables,
and a source term $\mb{\cal S}$ built from the $(1,0)$, $(0,1)$ quantities, see Eq.~(\ref{Einsep}).

As we explained in Section~\ref{framework},
this particular structure of the  $(1,1)$ equations  is quite
convenient in order to build an initial-boundary value problem and
solve it numerically by using time-domain methods.  The basic idea
is that given a numerical algorithm capable of evolving linear non-radial
perturbations, we can build an algorithm for our $(1,1)$ perturbations
by just adding source terms to the original algorithm.

The time evolution of non-radial perturbations of a static star
has been successfully analyzed by numerically integrating different systems
of perturbation equations~\cite{Allen:1998xj,Ruoff:2001ux,Nagar:2004ns}.
Taking into account the main features of our formulation, the scheme
introduced in~\cite{Nagar:2004ns,Nagar:2004pr} seems
to be  adequate for the purpose of implementing a numerical code to solve
our perturbation equations.
One of the main points in the scheme introduced in~\cite{Nagar:2004ns,Nagar:2004pr}
is the fact that the Hamiltonian constraint is not just an error
estimator for the evolution equations, as it is usually done in many free
evolution schemes.
In the scheme of~\cite{Nagar:2004ns,Nagar:2004pr}, the Hamiltonian constraint
is part of the system of equations and it is solved at every time
step for the perturbative quantity $k$, Eq.~(\ref{pk}).
This provides some control of the errors induced by constraint
violation.  As a consequence, the resulting numerical code~\cite{Nagar:2004ns}
is able to evolve non-radial perturbations for long times and is capable
to estimate the damping time and mode frequencies with an accuracy comparable to
frequency domain calculations.

The main idea of our present ongoing work~\cite{Passamonti:2005polar} on the
numerical solution of our perturbative equations is to follow the
scheme of~\cite{Nagar:2004ns,Nagar:2004pr}.
Taking into account our discussion in Section~\ref{framework} and above
about the general structure of the perturbative equations, in particular the same
differential structure of the perturbative equations at the orders $(0,1)$
and $(1,1)$, it is clear that this scheme is easily portable to our problem.
To that end, it is very important the fact that the Hamiltonian constraint
is solved for one of the perturbative quantities since at every time step we need
to evolve the equations for the $(0,1)$ and $(1,1)$ perturbations.  This means
that if we do not solve the Hamiltonian constraint, the errors accumulated from
constraint violation would be double than in a standard computation of
non-radial perturbations.   Therefore, the use of the scheme
of~\cite{Nagar:2004ns,Nagar:2004pr}
is a key ingredient in trying to obtain accurate long term evolutions.
We expect that the resulting numerical code would allow us to investigate
the non-linear effects of coupling.  In particular, we are interested in
looking for non-linear harmonics, possible resonances, parameter amplification, and/or changes in the damping time
of non-radial perturbations.

In the stellar interior, we evolve the (hyperbolic) equations for the metric
perturbation $\chi^{(1,1)}$ and for the fluid perturbation
$H^{(1,1)}$, which in some particular gauges coincides with the enthalpy
perturbation. The Hamiltonian constraint provides us the metric
perturbation $k^{(1,1)}$. Subsequently, all the other metric
$\psi^{(1,1)}$ and fluid ($\ga^{(1,1)}$, $\alpha^{(1,1)}$)
perturbations can be obtained from the perturbative equations
(\ref{psit}), (\ref{ktp}) and (\ref{psip}).

The wave equation for $\chi^{(1,1)}$ and the Hamiltonian constraint
are given by (\ref{chitt}) and (\ref{kpp}) respectively, while the
sound wave equation for $H^{(1,1)}$ has to be determined. We define
the fluid perturbation $H^{(1)}$ (see Appendix~\ref{AppGI} for a proof
of the gauge invariant character of this quantity) as
\begin{equation}
 H^{(1)} \equiv \frac{c_s^{2 (0)} \rho^{(0)}}{\rho^{(0)} +
p^{(0)}}   \om^{(1)}\,, \label{En_11}
\end{equation}
where the superscripts $(0)$ and $(1)$ have the meaning already explained
in Section~\ref{framework}.   The sound speed in the radially pulsating
spacetime can be split as follows
\begin{equation}
c_s^{2  {(0)}} = \bar{c}_s^2 + \la \,
\frac{d\bar{c}_s^2}{d\bar{\rho}}\,\delta\rho^{(1,0)}\,.
\end{equation}
In particular gauges, the Regge-Wheeler~\cite{Regge:1957} one for instance, the
gauge-invariant quantity $\om^{(1)}$ coincides with the
gauge dependent perturbation $\tilde{\om}^{(1)} $ [see
Eq.~(\ref{omgi})], and $H^{(1)}$ describes the enthalpy perturbation,
\begin{equation}
H^{(1)} \equiv \frac{\delta p^{(1)}}{\rho^{(0)} + p^{(0)}} \, ,
\label{En_11_RW}
\end{equation}
where $\delta p^{(1)}$ is defined by (\ref{pr}).  The wave equation
for $H^{(1,1)}$ is obtained as a linear combination of the time frame
derivative of equation (\ref{omegat}) and the spatial frame derivative
of (\ref{gammat}). After having introduced the equations (\ref{psit},
\ref{alphat}, \ref{gammat}, \ref{ktt}, \ref{kpp}) to reduce the number
of perturbative unknowns and the transformation
(\ref{En_11}), we have the following wave equation (written in the GSGM
formalism):
\begin{equation}
 -\ddot{H} +  c_s^2 H'' +  \mathcal{F}_{H} = 0 \, ,
\label{En_eq_GG}
\end{equation}
where $\mathcal{F}_{H}$ contains all the remaining terms (with
derivatives of lower order). The complete equation has been written
in Appendix \ref{AppSW11MG}. The wave equation (\ref{En_eq_GG}) is
valid in the GSGM framework for barotropic non-radial perturbations on
a time dependent background. In case of a static background, provided
the introduction of the background quantities
(\ref{unst}-\ref{mu_W_st}), it reduces to an equation well known in
the literature (see i.e. \cite{Allen:1998xj}, \cite{Ruoff:2001ux},
\cite{Nagar:2004ns}).

We can now write the perturbative equations for the stellar interior.
We consider instead of the perturbative quantity $\chi^{(1,1)}$, which diverges
like $r$ as we approach spatial infinity, the perturbation variable
$S^{(1,1)} = \chi^{(1,1)}/r$ which of course is well behaved at infinity.
This quantity satisfies the following \emph{gravitational wave} equation:
\beq
- S^{(1,1)}_{,tt} & + & e^{2 (\Phi - \Lambda)}
S^{(1,1)}_{,rr} + e^{2 (\Phi - \Lambda)} \left[
 \left( 5 \Phi_{,r}-\Lambda_{,r}
\right) S^{(1,1)}_{,r} + \frac{4}{r}
 \left( {\frac{1-{e^{2 \Lambda}}}{{r}^2}}+\Phi_{,r}^2+
 \frac{\Lambda_{,r}}{r} \right)k^{(1,1)} \right. \nn \\
 {}  & + & \left. \frac{1}{r} \left( \Phi_{,r} \left( 5 +4 \Phi_{,r}  r \right)
+3 \Lambda_{,r}+{\frac{2- \left( l(l+1) + 2 \right) e^{2\Lambda}}{r}} \right)
S^{(1,1)} \right]  =  e^{2  \Phi}{\cal S}_{S}\,. \label{GW11}
\eeq
where ${\cal S}_{S}$ denotes the source term for this wave equation.
The source terms in our $(1,1)$ perturbative equations have the following
pattern
\begin{equation}
{\cal S}^{(1,1)} = \sum_{I}{\cal C}^{(1,0)}_I {\cal Q}^{(0,1)}_I \,,
\label{pattern}
\end{equation}
which shows how the source terms introduce the coupling between radial
and non-radial perturbations in the $(1,1)$ equations.  In particular,
the source term in the gravitational-wave equation, ${\cal S}_S$
has the following form
\beq
{\cal S}_{S} & = & a_1  S_{,rr}^{(0,1)} + a_2  S_{,r}^{(0,1)} + a_3  S_{,t}^{(0,1)}
+ a_4  S^{(0,1)} + a_5  \left( \psi_{,r}^{(0,1)} - 2 e^{\Lambda-\Phi}
k_{,t}^{(0,1)} \right) + a_6  k^{(0,1)} + a_7  \psi^{(0,1)} \,, \label{Sgw11}
\eeq
where the coefficients $a_i$ are just linear combinations of
radial perturbations with coefficients constructed from background
quantities.  Their explicit form is given in Appendix~\ref{AppSources}.

The perturbative fluid variable $H^{(1,1)}$ also satisfies a wave equation,
but with a different propagation speed.  We call this equation the
\emph{sound  wave equation}.  It has the following form:
\begin{eqnarray}
- H^{(1,1)}_{,tt}   & + & \bar{c}_s^2 e^{2 \left(\Phi -\Lambda \right)}
H^{(1,1)}_{,rr}   +  e^{2 \left(\Phi-\Lambda\right)}
\left\{\left[ \left(   \frac{2}{r} + 2\Phi_{,r} - \Lambda_{,r} \right)
\bar{c}_s^2 - \Phi_{,r} \right] H^{(1,1)}_{,r}  \right. \nn \\
 & + &\left. \frac{1}{r} \left[ \left( 1 + 3\bar{c}_s^2\right)
 \left(\Lambda_{,r} + \Phi_{,r}  \right)-\bar{c}_s^2\frac{l(l+1)}{r}
 e^{2\Lambda} \right] H^{(1,1)} -  \frac{1-\bar{c}_s^2}{2}   \Phi_{,r}
\left[ \left(r S^{(1,1)}\right)_{,r} - k^{(1,1)}_{,r}\right] \right.  \nn \\
  & + & \left.
 \left[ -2 \Phi_{,r}^2 +\left[ \left(3\Phi_{,r} + \Lambda_{,r} \right) r
+ 1-  e^{2   \Lambda}  \right]  \frac{\bar{c}_s^2}{r^2}
\right]   \left( r   S^{(1,1)} + k^{(1,1)} \right)  \right\}
= e^{2 \Phi}   {\cal S}_{H}\,, \label{SW11}
\end{eqnarray}
and the source term can written as
\beq
{\cal S}_{H} & =  &   b_1  H^{(0,1)}_{,rr}
+ b_2  H^{(0,1)}_{,tr} + b_3 H^{(0,1)}_{,t} + b_4  H^{(0,1)}_{,r} +
b_5   H^{(0,1)}  + b_6  k^{(0,1)}_{,t} + b_7   r   S^{(0,1)}_{,t}  +
b_8 \left[ k^{(0,1)}_{,r} - \left(r S^{(0,1)}  \right)_{,r}\right] \nn \\
 {} & +  &  b_9  \left( r   S^{(0,1)} +  k^{(0,1)} \right) +
b_{10} \ga^{(0,1)}_{,r} + b_{11} \ga^{(0,1)} + b_{12} \psi^{(0,1)}_{,r}
+ b_{13}  \psi^{(0,1)} + b_{14} \alpha^{(0,1)} \,, \label{Ssw11}
\eeq
where the coefficients $b_i$ have the same structure are the
$a_i$ coefficients in~(\ref{Sgw11}).  Their explicit expressions
can be found in the Appendix~\ref{AppSources}.

For the last perturbative variable, the metric perturbation $k^{(1,1)}\,,$
we will use the \emph{Hamiltonian constraint} instead of an evolution equation.
After some calculations we get:
\beq
k^{(1,1)}_{,rr} & - &S^{(1,1)}_{,r}+ \left(\frac{2}{r} - \Lambda_{,r} \right)
k^{(1,1)}_{,r}+\frac{2}{r\bar{c}_s^2} \left(\Lambda_{,r}+\Phi_{,r}\right)
H^{(1,1)} + \frac{1}{r^2} \left[ \left( 1- l(l+1)\right) e^{2\Lambda}
+ 2\Lambda_{,r}r-1 \right] k^{(1,1)}   \nn \\
 & - & {} \frac{1}{2r} \left[ l(l+1) e^{2 \Lambda} + 4 - 4\Lambda_{,r} r
 \right] S^{(1,1)} = {\cal S}_{Hamil} \,, \label{Ham11}
\eeq
where ${\cal S}_{Hamil}$ is the source term for the Hamiltonian constraint.
As in the previous equations, it follows the pattern~(\ref{pattern}).
The precise form of ${\cal S}_{Hamil}$ is:
\beq
{\cal S}_{Hamil} & = &    c_1   \left( k^{(0,1)}_{,rr} - S^{(0,1)}_{,r} \right)
+ c_2 k^{(0,1)}_{,r}
+ c_3 k^{(0,1)}_{,t} + c_4     S^{(0,1)} + c_5    k^{(0,1)}
+ c_6 H^{(0,1)} + c_7 \psi^{(0,1)}_{,r}     \nn \\
  & + & {} c_8   \psi^{(0,1)} + c_9   \ga^{(0,1)}\,. \label{Sham11}
\eeq
The coefficients $c_{i}$, in the same way as the coefficients
$a_i$ and $b_i$ only contain radial perturbations
$g^{(1,0)}$ and quantities associated with the static background.
They are also given in Appendix~\ref{AppSources}.
It is worth to remark that the polar non-radial perturbation
equations on a static background are obtained from equations
(\ref{GW11},\ref{SW11},\ref{Ham11}) by discarding the source
terms and replacing all the $(1,1)$ perturbations with the
corresponding non-radial $(0,1)$.
The sources are determined from first order perturbations. The
radial perturbations from the equations
(\ref{eq:H10_ev}-\ref{eq:eta_cn}), and the non-radial perturbations
(described by the quantities $S^{(0,1)}\,,$ $k^{(0,1)}\,,$ and
$H^{(0,1)}$) from the first order analogous of the
above system (see~\cite{Nagar:2004ns}), and the equations
(\ref{psit}, \ref{ktp}, \ref{psip}) adapted to a static background
to get the $\psi^{(0,1)}$, $\ga^{(0,1)}$  and  $\alpha^{(0,1)}$.

The \emph{stellar exterior} is  described by a Schwarzschild
spacetime on which gravitational waves carry away some energy of the
stellar oscillations.   All fluid perturbations are not defined
outside the star and the radial perturbations vanish
because of Birkhoff's theorem.   Therefore,  the source terms in our
perturbation equations vanish. Only the metric perturbations
survive, and they satisfy the gravitational wave equation (\ref{GW11})
and the Hamiltonian constraint (\ref{Ham11}), which take the
following form:

\begin{eqnarray}
- S^{(1,1)}_{,tt}\! +e^{2(\Phi-\Lambda)} S^{(1,1)}_{,rr} +
 e^{2  \Phi} \left[\frac{6M}{r^2} S^{(1,1)}_{,r} - \left[ \frac{2M}{r^3}
\left(1 - \frac{2M}{r} e^{2\Lambda} \right) + \frac{l(l+1)}{r^2} \right]
S^{(1,1)} - \frac{4M}{r^4}\left(3 - \frac{M}{r} e^{2\Lambda}\right)
k^{(1,1)}  \right] = 0 \,, \nn \\
\end{eqnarray}
\begin{eqnarray}
e^{-2\Lambda}   \left( k^{(1,1)}_{,rr}  - S^{(1,1)}_{,r} \right)
 + \left( \frac{2}{r}  -\frac{3M}{r^2} \right) k^{(1,1)}_{,r} -
 \frac{l(l+1)}{r^2} k^{(1,1)} - \left(\frac{2}{r} - \frac{2M}{r^2}
 +\frac{l(l+1)}{2r}\right) S^{(1,1)} = 0 \,.
\end{eqnarray}
It is worth to mention that the above equations  coincide with
the equations for non-radial perturbations of a static stellar
background outside the star, as expected.

On the other hand, Zerilli showed that the even-parity perturbations
of a Schwarzschild background have just one degree of freedom, and
therefore can be described by just one variable, the Zerilli function, satisfying
a wave equation. At order $(1,1)$ the  Zerilli function can be built from the
two metric perturbations $S^{(1,1)}$ and $k^{(1,1)}$ and their derivatives, as
at first order~\cite{Moncrief:1974vm}, and is given by
\begin{equation}
 Z^{(1,1)} = \frac{2 r^2 e^{-2\Phi}}{(l+2)(l-1)r+6M}\left[r S^{(1,1)} +
 \frac{1}{2}\left(l\left(l+1\right) +\frac{2M}{r} \right)   e^{2\Phi}k^{(1,1)}
 - r k_{,r}^{(1,1)}  \right] \,.
\end{equation}
It satisfies the Zerilli equation~\cite{Zerilli:1970la,Zerilli:1970fj}
\begin{equation}
 - Z_{,tt}^{(1,1)} + e^{2 \left( \Phi -\Lambda \right)}Z_{,rr}^{(1,1)}
+ \frac{M}{r^2}   e^{2   \Phi}   Z_{,r}^{(1,1)} - V(r) Z^{(1,1)} =0\,,
\end{equation}
where $V(r)$ is the Zerilli potential~\cite{Zerilli:1970fj}:
\begin{equation}
V(r) = - \left(1-\frac{2M}{r}\right)\frac{n_{l}(n_{l}-2)^2 r^3
+6(n_{l}-2)^2 M r^2+ 36(n_{l} -2) M^2 r + 72M^3}{r^3[(n_{l}-2)r + 6M]^2} \,,
\end{equation}
where the quantity $n_{l} =l(l+1)$ has been
introduced to simplify the expression.

Finally, we can determine the power of the gravitational radiation emission
at infinity by using the following expression~\cite{Cunningham:1978cp}
\begin{equation}
\frac{d E}{d t} = \frac{1}{64 \pi}   \sum_{l\,, m} \,
\frac{\left( l + 2 \right)\, !}{\left(l-2\right)\, !} \, |\dot{Z}_{lm}|^2\,.
\end{equation}

\section{Boundary conditions}

In this Section we discuss the boundary conditions at the origin and
at the stellar surface for the $(1,1)$ perturbations describing the coupling
of radial and non-radial modes.  With regard to the outer boundary,
by locating it far enough we can use the well-known S\"ommerfeld outgoing boundary
conditions on our fields.

At the origin, the boundary conditions are just regularity conditions on the
perturbative fields, which can be obtained by a careful
analysis of the equations that they satisfy.
The analysis of Taylor expansions of the differential operators that appear in our
equations near the origin leads to the following behaviour for the non-radial
perturbations $\met^{(0,1)}$ and $\met^{(1,1)}$~\cite{Gundlach:1999bt}:
\beq
l & \ge & 0 \, ,   \quad S^{(1)} \sim  \, r^{l+1} \quad k^{(1)}
\sim  \, r^l \quad \psi^{(1)} \sim  \, r^{l+1}  \,, \\
l & \ge & 1 \, ,  \quad \ga^{(1)} \sim  \, r^{l-1} \quad H^{(1)}
\sim  \, r^l \, \quad  \alpha \sim \, r^l \,. \label{01_or_cond}
\eeq
By using the behaviour given by these expressions and also the behaviour
given in expressions~(\ref{bc_rad}) for the radial perturbations it  can be
proved the regularity of the source terms at the origin.

At the surface of the star, we have to consider a boundary condition
for the matter variable $H$, which vanishes in the spacetime region outside
the star.  For the metric perturbation variables $S$ and $k$, since they
must be continuous through the stellar surface, we can just use the
junction conditions at the surface to determine them.

To that end, let $\bar\Sigma$ be the surface of the static unperturbed star
(i.e. $r=R_s$).   The surface of the perturbed star can then be described
in the following way
\be
\Sigma\equiv\left\{x+\lambda\xi^{(1,0)}+\epsilon\xi^{(0,1)}
+\lambda\epsilon\xi^{(1,1)}\,:\,x\in\bar\Sigma\right\} \,,
\ee
where $\xi^{(i\,, j)}$ is a vector field that denotes the Lagrangian
displacement of a fluid element due to the action of perturbations of the
order $(i\,, j)$.
A physical requirement that follows from junction conditions is the
vanishing of the unperturbed pressure $\bar{p}$ at the unperturbed
surface $\bar\Sigma$.  In the same way, the corresponding boundary condition
for the perturbed spacetime is the vanishing of the total pressure $\bar{p} +\lambda
\delta p^{(1,0)}+\epsilon\delta p^{(0,1)}+\lambda\epsilon\delta p^{(1,1)}$ at
the perturbed surface $\Sigma$.  This condition turns out to be equivalent to
the vanishing of the Lagrangian pressure perturbations on $\bar\Sigma$, the unperturbed
surface, at every order.  The Lagrangian pressure perturbations are given by:
\beq
\Delta \, p^{(1,0)}  &=& \delta \, p^{(1,0)}  + \pounds_{\xi_{(1,0)}} \, \bar{p} \,,
\label{LPp10} \\
\Delta \, p^{(0,1)}  &=& \delta \, p^{(0,1)}  + \pounds_{\xi_{(0,1)}} \, \bar{p} \,,
\label{LPp01} \\
\Delta \, p^{(1,1)}  &=&
\delta \, p^{(1,1)}  + \left(\pounds_{\xi_{(1,1)}} +\frac{1}{2}
\left\{\pounds_{\xi_{(1,0)}}\,,\,\pounds_{\xi_{(0,1)}}
\right\} \right) \bar{p} + \pounds_{\xi_{(0,1)}}\delta p^{(1,0)}
+ \pounds_{\xi_{(1,0)}}\delta p^{(0,1)} \nn\\
&=& \delta \, p^{(1,1)}  + \left( \pounds_{\xi_{(1,1)}}
-\frac{1}{2}
\left\{\pounds_{\xi_{(1,0)}}\,,\,\pounds_{\xi_{(0,1)}} \right\}
\right) \bar{p}\,, \label{p11atsf}
\eeq
where $\delta$ and $\Delta$ denote the Eulerian and Lagrangian
perturbations respectively, and we have used the lower order
boundary conditions $\Delta p^{(1,0)}=\Delta p^{(0,1)}=0$ in order
to simplify the condition (\ref{p11atsf}).

{}From this analysis we can conclude that the boundary conditions for the fluid
perturbations
are described by the set of expressions given in~(\ref{LPp10}-\ref{p11atsf}).
However, in practice, in many applications of first order perturbation theory,
{\em dynamical
boundary conditions} either for density or enthalpy perturbations have been considered.
This alternative boundary conditions follow from the analysis of the time
derivative of the condition (\ref{LPp01}) (see~\cite{Allen:1998xj} for more
details).
In our current development of the numerical implementation of the perturbative
equations we are considering both types of boundary conditions with the
perspective of analyzing which type works best for our formulation.

Finally, the junction conditions for the metric perturbations can be determined
by imposing continuity of first and second fundamental differential
forms and their perturbations at
the surface~\cite{Darmois:1927gd,Lichnerowicz:1971al,Obrien:1952bs,Israel:1966nc}.
The explicit form of these conditions has been presented in~\cite{Martin-Garcia:2000ze}
for first order perturbations of a time-dependent stellar background.
Alternatively, one may use the ``extraction formulas"~\cite{Martin-Garcia:2000ze}
that relate the Zerilli function with metric perturbations at the stellar
boundary.

\section{Remarks and Conclusions}

Non linearity is the rule rather than the exception in dynamical phenomena in all
branches of physics. The modeling of compact objects such as neutron stars and
supernov\ae \ core must ultimately be rooted in general relativity (or some of its
generalisations),  where non-linearity  represents a fundamental physical
character of the theory, i.e.\ the self-interaction of the gravitational field,
and not  just corrections to
an underlying linear modeling of gravitational phenomena.
In relativistic theories of gravity, gravitational radiation is the typical outcome
of dynamical phases in the life of sources such as binary systems and supernov\ae,
and  major experimental efforts are currently under way to detect this most elusive
prediction of Einstein gravity for the first time.  This will eventually  lead to the
development of a whole new  branch of  astronomy, based on observing gravitational
radiation, much in the same way it  has been in the past for x-rays and other parts
of the electromagnetic spectrum outside the visible band.
In this context the accurate
theoretical modeling of sources is crucial to the final end of providing templates
in this game of looking for a  needle - the signal - in the haystack, the noise.
While ultimately a full numerical relativity description of gravitational wave
sources is needed to model the most non-linear dynamical phases, much interesting
physics can be understood by using approximate methods. Furthermore, a semi-analytical
approach typically  helps to shed light on the physical processes, thus complementing
the numerical work.

Relativistic perturbation theory is ideal for those cases where a known solution of
the field equations is explicitly known, as for black holes, or can easily be obtained,
as is the case for compact stars. An advantage of the relativistic perturbative
approach is that it directly incorporates gravitational waves. For smaller
perturbations, linear theory suffices. If one wants to consider mildly non-linear
oscillations of a compact object, second order effects and mode coupling have to be
taken into account. For black holes, many studies already exists in this direction
(see e.g.\  \cite{Gleiser:1995gx,1999bhgr.conf..351P,Campanelli:1998jv}).
In the case of neutron stars,  while linear perturbations and instabilities have
been studied for long time~\cite{Andersson:2002ch,Kokkotas:2002ng}, relatively
little is known of non-linear dynamical effects, mostly through numerical studies
(see e.g.\
\cite{Sperhake:2001xi,2002PhRvD..65b4001S,2002ApJ...571..435M,2002PhRvD..65h4039L,2003ApJ...591.1129A,Font:2001ew,Stergioulas:2003ep}).
A second order  perturbative approach is therefore timely and  may help to
understand known problems and even reveal a new  phenomenology.

In this paper we have developed the relativistic formalism to study  a particular
second order effect, the coupling of radial and non-radial first order  perturbations
of a compact spherical star. From a mathematical point of view it is very convenient to
treat the two sets of  perturbations, radial and
non-radial,  as separately parametrized, using the multi-parameter
perturbative formalism developed in~\cite{Bruni:2002sm,Sopuerta:2003rg}.
Then    we have considered the expansion of the metric, the energy-momentum tensor and
Einstein equations in terms of two parameters  $\lambda$ and $\epsilon$, where
$\lambda$ parametrizes the radial modes, $\epsilon$ the non-radial
perturbations, and the $\lambda\epsilon$ terms describe the coupling. This approach
provides a well-defined framework to   consider the gauge dependence of
perturbations. In this mathematical context we have imported the formalism of
Gundlach and Mart\'{\i}n Garc\'{\i}a~\cite{Gundlach:1999bt,Martin-Garcia:2000ze} and
 Gerlach and Sengupta~\cite{Gerlach:1979rw}, describing gauge-invariant perturbations
 of a general time--dependent spherical
spacetime, expanding the latter in a static background and a radial perturbation.
Fixing the gauge for radial perturbations allows us to: {\it i)} use
the GSGM gauge-invariant non-radial $\epsilon$ variables on the static background;
{\it ii)} define new second order $\lambda\epsilon$ variables, describing the
non-linear coupling of the the radial and non-radial linear perturbations, that
are also gauge-invariant at the $\lambda\epsilon$ second order.  This higher order
gauge invariance, attained by partially fixing the gauge at first order,  is similar
to that considered for example in~\cite{Cunningham:1980cp} and~\cite{Gleiser:1995gx}.
In our case however we use  a 2-parameter $\lambda - \epsilon$
expansion~\cite{Bruni:2002sm,Sopuerta:2003rg}, so that  we only need to fix the
gauge for radial perturbations. Assuming a barotropic perfect fluid, we have derived
the evolution and constraint equations for our variables, in particular those for
the coupling terms of order   $\lambda\epsilon$, focusing on polar perturbations.
We leave for future studies the implementation of more realistic equations of state, such as the non-isentropic one used in \cite{Font:2001ew,Stergioulas:2003ep}.
As expected, in the interior the $\lambda\epsilon$ variables
satisfy inhomogeneous linear equations where the homogeneous part is governed by
the same linear operator acting on the first order $\epsilon$ non-radial perturbations,
while the source terms are quadratic and made of products of $\lambda$ and $\epsilon$
terms. In the exterior there is no direct coupling, and the whole dynamics is embodied
in the $\lambda\epsilon$ order Zerilli function. Thus the effect of the coupling is
transmitted from the interior to the exterior through the junction conditions at the surface of the star.
Finally, we have given a brief discussion of the boundary conditions, focusing on
those on the surface. These are typically expressed in terms of the
Lagrangian pressure perturbation, therefore we have defined a $\lambda\epsilon$ second
order Lagrangian displacement  and a corresponding $\lambda\epsilon$ Lagrangian pressure
perturbation, appropriately related to the Eulerian perturbation. Thanks  to the
vanishing of the first order $\lambda$ and $\epsilon$ Lagrangian pressure perturbations
on the surface, this relation turns out to be linear. 

Work is currently under way, numerically  implementing the
formalism
presented here, in order to provide a first analysis of the possible effects of the
coupling between radial and polar non-radial perturbations~\cite{Passamonti:2005polar}. Some of these effects  are easily
anticipated for  the case of axial oscillations. These are decoupled from
fluid perturbations at first order,  but are driven by  the radial pulsations at the $\lambda\epsilon$
order~\cite{Passamonti:2004axial}. Eventually these studies may possibly even lead to discover new unexpected effects
of mode coupling. Surely we expect to find non-linear harmonics arising from the radial non-radial coupling, similar to those between various radial modes found in~\cite{Zanotti:2004kp} for tori around black holes, a prediction that appears to be confirmed by a numerical relativity study of neutron stars in the conformally-flat spacetime approximation~\cite{Dimmelmeier:2004prep}.



\[ \]
{\bf Acknowledgements:}
This work has been partially
supported by the EU Network Programme (Research Training Network
contract HPRN-CT-2000-00137).
CFS has been supported by EPSRC in the first stages of this work and presently
is partially supported by NSF grants PHY-9800973 and PHY-0114375.
The authors wish to thank Nick Stergioulas for carefully reading the manuscript, and Nils Andersson, Christian Cherubini, Valeria Ferrari, Kostas Glampedakis, Kostas Kokkotas, Giovanni Miniutti, and Alessandro Nagar for fruitful discussions.

\appendix

\section{Source terms for the $(1,1)$ perturbation equations} \label{AppSources}

\noindent In this Section we give the expressions of the coefficients
appearing in  the source terms of the
equations (\ref{GW11}-\ref{Sham11}) for the $(1,1)$ perturbations.

In the case of the \emph{gravitational wave} equation~(\ref{Sgw11})
the source term has the form:
\beq
{\cal S}_{S} & =  &   a_1  S_{,rr}^{(0,1)} + a_2 S_{,r}^{(0,1)}
+ a_3 S_{,t}^{(0,1)} + a_4 S^{(0,1)} + a_5 \left( \psi_{,r}^{(0,1)}
- 2 e^{\Lambda -\Phi} k_{,t}^{(0,1)} \right) + a_6 k^{(0,1)}  +
a_7 \psi^{(0,1)} \,,
\eeq
where the coefficients $a_i$ are given by
\beq
a_1 =  2\left( r S^{(1,0)}- \eta^{(1,0)}\right) e^{-2\Lambda} \,,
\eeq
\beq
a_2 = \left[ 2\left(\Lambda_{,r}-5\Phi_{,r}\right)\eta^{(1,0)}
- \left(\left( \Lambda_{,r}-5\Phi_{,r} \right) r+3 \right) S^{(1,0)}  -
\left( \Lambda_{,r}+\Phi_{,r} \right)\left(5-\bar{c}_s^{\,-2} \right)
H^{(1,0)}\right]{e^{-2\,\Lambda}} - 4\ga_{,t}^{(1,0)} e^{-\Phi-\Lambda} \,,
\eeq
\beq
a_3 = - 4\left( \Lambda_{,r}+\Phi_{,r}\right)\ga^{(1,0)}
 e^{-\Lambda-\Phi} - e^{-2\Phi} \eta^{(1,0)}_{,t}  + \frac{2}{r}
 \left( r\ga^{(1,0)} e^{-\Phi} \right)_{,r} e^{-\Lambda}    \,,
\eeq
\beq
a_4 & = & -\left\{\frac{4}{r}\left(1+2r\Phi_{,r}\right)\ga^{(1,0)}_{,t}
e^{-\Phi+\Lambda} + 2\left[2\Phi_{,r} \left( \frac{1}{r} + 2\Phi_{,r}
\right) +\frac{2-\left(l(l+1)+2\right)e^{2\Lambda}}{r^2}+3\frac{\Lambda_{,r}
+\Phi_{,r}}{r} \right]\eta^{(1,0)} \right. \nn \\
{} & - &  \left. \left[\Phi_{,r}+3\Lambda_{,r}+\frac{3-\left(l(l+1)+2\right)
e^{2\Lambda}}{r}\right] S^{(1,0)}
+ \left( \Lambda_{,r}+\Phi_{,r} \right) \left[\left(5+\frac{3}{\bar{c}_s^2}\right)
\frac{1}{r} + 8\Phi_{,r}\right]H^{(1,0)}\right\}e^{-2\Lambda} \,,
\eeq
\beq
a_5 = - 2\left[\left(\frac{e^\Phi}{r}\ga^{(1,0)}\right)_{,r}
-\left(\Lambda_{,r}+\Phi_{,r}\right)\left(\frac{e^\Phi}{r}\ga^{(1,0)}\right)
\right]e^{-2\Lambda-\Phi}\,,
\eeq
\beq
a_6 & = & -\left[\frac{2}{r^2}\left(-5+2r(\Phi_{,r}-\Lambda_{,r})+2e^{2\Lambda}
\right) S^{(1,0)} +\frac{8}{r^3}\left(1+r^2\Phi_{,r}^2+r\Lambda_{,r}-
e^{2\Lambda}\right)\eta^{(1,0)} \right. \nn \\
{} & + & \left. \frac{4}{r^2} \left(\Lambda_{,r}+\Phi_{,r}\right)
\left(2r\Phi_{,r}+\frac{1}{\bar{c}_s^2}\right)H^{(1,0)}
+\frac{8}{r}\Phi_{,r}\ga^{(1,0)}_{,t}e^{-\Phi+\Lambda}\right]
e^{- 2\Lambda} \,,
\eeq
\beq
a_7 & = & -\frac{2}{r}\left\{\left(1-\bar{c}_s^2\right)r \left(
\frac{e^\Phi}{r}\ga^{(1,0)}\right)_{,rr}\!\!\! +
\left( \frac{e^{\Phi}}{r}\ga^{(1,0)}\right)_{,r}
\left[r\left(\Phi_{,r}-2\Lambda_{,r}\right) + \left(2\Lambda_{,r}r
+\Phi_{,r}r-4\right)\bar{c}_s^2 \right.
\right. \nn \\
{} & + & \left. \frac{\Phi_{,r}}{4\pi} \frac{\Lambda_{,r}+
\Phi_{,r}}{\bar{c}_s^2} \frac{d\bar{c}_s^2}{d\bar\rho}e^{-2\Lambda}
\right] +  \left[\left(2-2r\Phi_{,r}
-3r\Lambda_{,r}\right)\Phi_{,r}-\left(1+r\Lambda_{,r}\right)\Lambda_{,r}
+ \left(r\left(\Lambda_{,r}^2- \Phi_{,r}^2\right)
+2\Phi_{,r} + 5\Lambda_{,r}\right)\bar{c}_s^2 \right.  \nn \\
{} & + & \left. \left. \frac{\Lambda_{,r}+\Phi_{,r}}
{\bar{c}_s^2}\Phi_{,r}\left[r+\frac{e^{-2\Lambda}}{4\pi r}
\left(3-\left(\Lambda_{,r}+\Phi_{,r}\right)r \right)\frac{d\bar{c}_s^2}
{d\bar\rho} \right] \frac{e^{\Phi}}{r}\ga^{(1,0)}\right]
 \right\} e^{-\Phi -2\Lambda} \,.
\eeq

For the \emph{sound wave} equation the source term is given by (\ref{Ssw11}):
\beq
{\cal S}_H & = & b_1 H^{(0,1)}_{,rr}+b_2 H^{(0,1)}_{,tr}
+ b_3 H^{(0,1)}_{,t} + b_4 H^{(0,1)}_{,r} + b_5 H^{(0,1)}
+ b_6 k^{(0,1)}_{,t} + b_7 r S^{(0,1)}_{,t} + b_8\left[k^{(0,1)}_{,r}
-\left(r S^{(0,1)}\right)_{,r}\right] \nn \\
& + & b_9\left(r S^{(0,1)} + k^{(0,1)}\right) + b_{10} \ga^{(0,1)}_{,r}
+ b_{11} \ga^{(0,1)} + b_{12} \psi^{(0,1)}_{,r} +
b_{13}\psi^{(0,1)} + b_{14}\alpha^{(0,1)} \,,
\eeq
and the expression of the coefficients $b_i$ is the following
\beq
b_1 = -\left[2\left(\eta^{(1,0)}-rS^{(1,0)}\right)\bar{c}_s^2+
\frac{e^{-2\Lambda}}{4\pi r} \frac{\Lambda_{,r}+\Phi_{,r}}{\bar{c}_s^2}
\frac{d\bar{c}_s^2}{d\bar\rho}H^{(1,0)} \right]e^{-2\Lambda}\,,
\eeq
\beq
b_2 = 2\left(1-\bar{c}_s^2\right) e^{-\Phi-\Lambda}\ga^{(1,0)} \,,
\eeq
\beq
b_3 & = & - \left\{\frac{e^{-\Lambda}}{r^3} \left(-\frac{e^{-2\Lambda}}{2\pi}
\frac{\Lambda_{,r}+\Phi_{,r}}{\bar{c}_s^2} \frac{d\bar{c}_s^2}
{d\bar\rho} + \frac{2\bar{c}_s^2\bar\rho - \bar{p}}{\bar\rho}r\right )
\left(r^2\ga^{(1,0)} e^{\Phi}\right)_{,r} \right. \nn \\
& + & \left.  \left(\Lambda_{,r}+\Phi_{,r}\right)
\left(\frac{e^{-2\Lambda}}{2\pi r}\frac{\Lambda_{,r}+\Phi_{,r}}{\bar{c}_s^2}
\frac{d\bar{c}_s^2}{d\bar\rho}-\bar{c}_s^2+1+
\frac{\bar{p}}{\bar\rho}\right)\ga^{(1,0)}
e^{\Phi-\Lambda}+\eta^{(1,0)}_{,t}\right\} e^{-2\Phi}\,,
\eeq
\beq
b_4 & = & \frac{1}{4\pi r} \left[\Lambda_{,r}\left(\Lambda_{,r}-
\frac{1}{r}\right)-\left(\frac{1}{r}+\Phi_{,r}\right)\left(2\Phi_{,r}+
\Lambda_{,r}\right)\right] \frac{1}{\bar{c}_s^2}
\frac{d\bar{c}_s^2}{d\bar\rho}H^{(1,0)} e^{-4\Lambda} \nn \\
& - & \left\{2\left(1-\bar{c}_s^2\right ) H^{(1,0)}_{,r}
+2\left[\left( 2\Phi_{,r}-\Lambda_{,r}+\frac{2}{r}\right) \bar{c}_s^2
-\Phi_{,r}\right] \eta^{(1,0)} \right. \nn \\
& + & \left. \left[ 3\Phi_{,r}+\frac{1}{2r}+\left(\Lambda_{,r} -
4\Phi_{,r} - \frac{7}{2r}\right)\bar{c}_s^2\right] r S^{(1,0)}\right\}
{e^{-2\Lambda}} \,,
\eeq
\beq
b_5 & = &  -\left\{\left[\frac{e^{-2\Lambda}}{4\pi r^2}
\left(1-{e^{2\Lambda}}+\Phi_{,r}\Lambda_{,r}r^2+ \left(\Lambda_{,r}
+\frac{5}{2}\Phi_{,r}\right)r\right)
\frac{\Lambda_{,r}+\Phi_{,r}}{\bar{c}_s^2}\frac{d\bar{c}_s^2}
{d\bar\rho} - \left(\Lambda_{,r}+\Phi_{,r}\right) \left(1+3\bar{c}_s^2\right)
\right. \right. \nn \\
& + & \left. \left. \frac{l(l+1)}{r}\bar{c}_s^2 e^{2\Lambda}\right] S^{(1,0)} +
\frac{\Lambda_{,r}+\Phi_{,r}}{\bar{c}_s^2 r}\left[\frac{1}{4\pi r}
\left( \left(6\bar{c}_s^2+1+ r\Phi_{,r}\right)
\frac{\Lambda_{,r}+\Phi_{,r}}{\bar{c}_s^2}{e^{-2\Lambda}}
-\frac{l(l+1)}{r}\right )\frac{d\bar{c}_s^2}{d\bar\rho}  \nn \right.\right.\\
& + & \left. \left. 3\bar{c}_s^4+4\bar{c}_s^2+1\right] H^{(1,0)} +
\frac{e^{-2\Lambda}}{4\pi r} \frac{\Lambda_{,r}+\Phi_{,r}}{\bar{c}_s^2}
\frac{d\bar{c}_s^2}{d\bar\rho}\left[ \left(
\frac{2}{r} - \Lambda_{,r}+ 2\Phi_{,r} \right)
 H^{(1,0)}_{,r} + H^{(1,0)}_{,rr} \right] \right. \nn    \\
& + & \left. \frac{2}{r}\left[ \left(\Lambda_{,r}+\Phi_{,r}\right )
\left(1+3\bar{c}_s^2\right) -\frac{l(l+1)}{r}\bar{c}_s^2
e^{2\Lambda} \right] \eta^{(1,0)}\right\}{e^{-2\Lambda}} \,,
\eeq
\beq
b_6 = \left\{\left[ \left(\Lambda_{,r}+\Phi_{,r} \right)\left(1+\bar{c}_s^2
-{\frac{\bar{p}}{\bar\rho}} \right)\bar{c}_s^2-\left(1-\bar{c}_s^2\right )
\Phi_{,r}\right]
\ga^{(1,0)} e^{\Phi} - \frac{\bar{c}_s^2}{r^2} \left(1+\bar{c}_s^2-
\frac{\bar{p}}{\bar\rho}\right)
\left(r^2\ga^{(1,0)} e^{\Phi} \right)_{,r} \right\}e^{-2\Phi-\Lambda} \,,
\eeq
\beq
b_7 = \left\{\left[ \left( \Lambda_{,r}+\frac{2}{r} \right )\bar{c}_s^2 +
\frac{\Phi_{,r}}{2}\left(1+\bar{c}_s^2\right) \right]\ga^{(1,0)}e^{\Phi} -
\frac{\bar{c}_s^2}{r^2} \left( r^2 \ga^{(1,0)}e^{\Phi} \right)_{,r}
 \right\} {e^{-2\Phi-\Lambda}} \,,
\eeq
\beq
b_8 =  -\left[  \frac{\Phi_{,r}}{8\pi r}\frac{\Lambda_{,r}+\Phi_{,r}}
{\bar{c}_s^2}\frac{d\bar{c}_s^2}{d\bar\rho} H^{(1,0)}
e^{-2\Lambda} + \left(1-\bar{c}_s^2 \right)
\left( \Phi_{,r} \eta^{(1,0)} + \frac{1}{2}H^{(1,0)}_{,r} \right )\right]
e^{-2\Lambda} \,,
\eeq
\beq
b_9 & = & -\left\{\left[ \frac{2}{r^2}\left(1-{e^{2\Lambda}}+3r\Phi_{,r}
+r\Lambda_{,r}\right)\bar{c}_s^2 -4{\Phi_{,r}}^2 \right] \eta^{(1,0)}
\nn \right.   \\
& + & \left. \left[\left[3\frac{e^{2\Lambda} -1}{r}  - 2\Phi_{,r}
\left( 4 + r \Lambda_{,r} \right) - 3\Lambda_{,r} \right]\bar{c}_s^2 +4 r
\Phi_{,r}^2\right] S^{(1,0)} \nn \right.   \\
& - & \left. \frac{\Lambda_{,r}+\Phi_{,r}}{r} \left[ 1+ 2\Phi_{,r} r
+3\bar{c}_s^2 - \frac{e^{-2\Lambda}}{4\pi r^2} \frac{1}{\bar{c}_s^2}
\frac{d\bar{c}_s^2}{d\bar\rho} \left(\left(3 \Phi_{,r}
+\Lambda_{,r}\right)r+ 1 - e^{2\Lambda} \right)\right]
H^{(1,0)} \nn \right.   \\
& + & {} + \left. 2\left[ 2\Phi_{,r}+\left( \Lambda_{,r}-2\Phi_{,r}-
\frac{2}{r}\right ) \bar{c}_s^2 \right] H^{(1,0)}_{,r} - 2\bar{c}_s^2
H^{(1,0)}_{,rr}\right\} e^{-2\,\Lambda} \,,
\eeq
\beq
b_{10} & = &  \frac{2   \bar{c}_s^2}{r^2}
\left\{\left[  \left( 1-{\frac{\bar{p}}{2\bar\rho}}  \right) \left
(\Lambda_{,r}+\Phi_{,r}\right ) +   \frac{2}{r} \right]
  \left( r^2   \ga^{(1,0)}   e^{\Phi}  \right) - \left(  1
-{\frac{p}{2   \rho}}\right)     \left( r^2   \ga^{(1,0)}
  e^{\Phi}  \right)_{,r}
\right\}   e^{-\Phi-2 \Lambda} \,,
\eeq
\beq
b_{11} &  = &  - \frac{1}{r^3}   \left\{2  r
\bar{c}_s^2 \left(1-\bar{c}_s^2\right )   \left( r^2
  \ga^{(1,0)}   e^{\Phi}  \right)_{,rr} + \left[
\frac{\Phi_{,r}}{2   \pi}   \left(\Lambda_{,r}+\Phi_{,r}\right )
\frac{d\bar{c}_s^2}{d\bar\rho}  {e^{-2 \Lambda}}
+2 \left(\Phi_{,r}+2 \Lambda_{,r}+\frac{2}{r} \right )r   \bar{c}_s^4
 \right. \right. \nn \\
& - & \left. \left. \left( 4 \left(1+r   \Lambda_{,r}
\right) + {\frac{\bar{p}}{\bar\rho} \left(2+ r   \Phi_{,r} \right )
}\right )\bar{c}_s^2 - \left(2-{\frac{\bar{p}}{\bar\rho}}\right ) r
\Phi_{,r} \right] \left( r^2   \ga^{(1,0)}   e^{\Phi}
\right)_{,r} +
 \left[ -2 \left
(\Lambda_{,r}+\Phi_{,r}\right )\left(1+\left(\Phi_{,r}-
\Lambda_{,r}\right )r\right ) \bar{c}_s^4 \right.
\right. \nn \\
& + & \left. \left. \left(  \frac{5}{r}    -\frac{\Phi_{,r}}{2}
\left( 2 \Phi_{,r}r+7+8 r\Lambda_{,r}\right) - 2 \Lambda_{,r}
\left(r\Lambda_{,r}-1\right )+\frac{\bar{p}}{\bar\rho}\frac{\Phi_{,r}
r \left(2 \Phi_{,r}r+5\right )+2}{2   r}
-\frac{e^{2 \Lambda}}{2  r}   \left(1+{\frac{\bar{p}}{\bar\rho}}\right)
\left( 2+\Phi_{,r}r\right )   \right )\bar{c}_s^2  \right.  \right. \nn \\
& + & \left. \left. \left(\left(4-{\frac{\bar{p}}{\bar\rho}}\right
)\left(\Lambda_{,r}+\Phi_{,r}\right )r+4\right )\Phi_{,r} -
\frac{e^{-2\Lambda}}{2   \pi}  \left(\Lambda_{,r}+\Phi_{,r}\right )^2
\Phi_{,r} \frac{d\bar{c}_s^2}{d\bar\rho}\right]
\left( r^2 \ga^{(1,0)} e^{\Phi}  \right) \right\} e^{-\Phi-2 \Lambda}\,,
\eeq
\beq
b_{12} =  \frac{\bar{c}_s^2}{2   r^2}
  \left\{\left[  \left
(\Lambda_{,r}+\Phi_{,r}\right )\left
(1+\bar{c}_s^2\right )+\frac{4}{r} \right]   \left( r^2
\ga^{(1,0)}   e^{\Phi} \right) -  \left
(1+\bar{c}_s^2\right ) \left( r^2   \ga^{(1,0)}
e^{\Phi}  \right)_{,r} \right\} e^{-\Phi-2 \Lambda}    \,,
\eeq
\beq
b_{13} & = &  - \frac{1}{r^2}   \left\{\bar{c}_s^2
\left( 1 + \bar{c}_s^2 \right)   \left( r^2   \ga^{(1,0)}
  e^{\Phi}  \right)_{,rr} \!\!+
\frac{e^{-2 \Lambda}}{4   \pi   r}    \left(\Lambda_{,r}+\Phi_{,r} \right )
\Phi_{,r}   \left[  \left(\Lambda_{,r}+\Phi_{,r} \right )
\left( r^2\ga^{(1,0)} e^{\Phi} \right) - \left( r^2\ga^{(1,0)}
e^{\Phi} \right)_{,r} \right]  \frac{d\bar{c}_s^2}{d\bar\rho} \right. \nn \\
& - & \left.
\left[ \left(  2    \Lambda_{,r}+\Phi_{,r} \right)   \bar{c}_s^4 +
\left(\frac{2}{r} + 2\Lambda_{,r} + 3 \Phi_{,r} + \frac{\bar{p}}{\bar\rho}
\left( \frac{1}{r} - \frac{\Phi_{,r}}{2}\right) \right) \bar{c}_s^2 +
\left(2 -\frac{\bar{p}}{2\bar\rho} \right)\Phi_{,r}\right] \left( r^2 \ga^{(1,0)}
e^{\Phi} \right)_{,r}  \right. \nn \\
& + &  \left. \left[ \left( \frac{4}{r} + \frac{\Lambda_{,r}+\Phi_{,r}}{2}
\left( 6 - \frac{\bar{p}}{\bar\rho} \right) \right)\Phi_{,r}  -
\left( \Lambda_{,r}+\Phi_{,r} \right)
\left( \Lambda_{,r} - \Phi_{,r} + \frac{1}{r} \right)
\bar{c}_s^4 \right. \right. \nn \\
& + & \left. \left.
\left[ \frac{1}{2} \left( \Phi_{,r} + \frac{1}{2   r}
\right) \left( 6 \Lambda_{,r} + 7  \Phi_{,r} -
\frac{\bar{p}}{\bar\rho} \left( \Phi_{,r} -\frac{2}{r} \right)  \right)
+ \frac{2}{r^2} - \left( \Lambda_{,r} - \frac{1}{2   r}
\right)
\left( \Lambda_{,r} + \frac{1}{r} \right) \right. \right. \right. \nn \\
& + & \left. \left. \left.
\frac{1}{4   r^2}   \left( 1 + \frac{\bar{p}}{\bar\rho} \right) \left( r
\Phi_{,r} -2 \right) e^{2\Lambda}\right]   \bar{c}_s^2\right]r^2
\ga^{(1,0)}   e^{\Phi}\right\} e^{-\Phi - 2\Lambda}\,,
\eeq
\beq
b_{14} & = & -  \frac{\bar{c}_s^2}{r^3}
\left\{\left[  2  \left(\Lambda_{,r}+\Phi_{,r} \right)
\left(1+\bar{c}_s^2-{\frac{\bar{p}}{\bar\rho}} \right ) +
\frac{\bar{p}}{\bar\rho}    \frac{l \left( l + 1\right)}{r}
e^{2 \Lambda}\right]  \left[  \left( r^2   \ga^{(1,0)}
e^{\Phi}  \right)_{,r} -\left(\Lambda_{,r}+\Phi_{,r}\right)r^2
\ga^{(1,0)}   e^{\Phi}  \right]e^{-2 \Lambda}  \right. \nn \\
& - & \left. 2 l \left( l + 1\right)       \ga^{(1,0)}
e^{\Phi} \right\}  e^{-\Lambda- \Phi}\,.
\eeq

Finally, in the case of the \emph{Hamiltonian constraint}, the equation
we have considered for $k^{(1,1)}$, the source term is given by~(\ref{Sham11}):
\beq
{\cal S}_{Hamil} & = & c_1\left( k^{(0,1)}_{,rr} - S^{(0,1)}_{,r}\right)
+ c_2 k^{(0,1)}_{,r} + c_3 k^{(0,1)}_{,t} + c_4 S^{(0,1)}
+ c_5 k^{(0,1)} + c_6 H^{(0,1)} + c_7 \psi^{(0,1)}_{,r} +  {} \nn \\
& + & c_8 \psi^{(0,1)} + c_9 \ga^{(0,1)}  \,,
\eeq
where the coefficients $c_i$ are given by
\beq
c_1 & = &  r  S^{(1,0)} \,, \\
c_2 & = & \left( \frac{3}{2}  S^{(1,0)} +  \frac{\Lambda_{,r}+
\Phi_{,r}}{\bar{c}_s^2}    H^{(1,0)} \right) \,, \\
c_3 & = &  - \left(\Lambda_{,r}+\Phi_{,r}\right)e^{\Lambda-\Phi}\ga^{(1,0)}  \,, \\
c_4 & = &  - \left(S^{(1,0)} + 2\frac{\Lambda_{,r}+\Phi_{,r}}{\bar{c}_s^2}
H^{(1,0)} \right)\,,  \\
c_5 & = &  -  \frac{2}{r} \frac{\Lambda_{,r}+\Phi_{,r}}{\bar{c}_s^2}H^{(1,0)}  \,, \\
c_6 & = &  - \frac{2}{r} \frac{\Lambda_{,r}+ \Phi_{,r}}{\bar{c}_s^2}\left[ 1+
\frac{1}{\bar{c}_s^2} - \frac{e^{-2 \Lambda}}{4\pi r}\frac{\Lambda_{,r}+
\Phi_{,r}}{\bar{c}_s^4}\frac{d\bar{c}_s^2}{d\bar\rho}\right] H^{(1,0)} \,,   \\
c_7 & = & \frac{2}{r}\ga^{(1,0)}  \,,  \\
c_8 & = & \frac{1}{r^2}\left[\left(2-4\Lambda_{,r}r + l\left(l+1\right)
{e^{2\Lambda}}\right)\ga^{(1,0)} + 2 r\ga_{,r}^{(1,0)} \right] \,,  \\
c_9 & = &  -\frac{4}{r}\left(\Lambda_{,r}+\Phi_{,r}\right)\ga^{(1,0)} \,. \label{ccc}
\eeq

\section{Sound Wave Equation} \label{AppSW11MG}

\noindent We write here the complete sound wave equation
(\ref{En_eq_GG}) of a generic barotropic fluid perturbation on a
time-dependent background in terms of the GSGM quantities,
\begin{eqnarray}
& - & \ddot{H} + c_s^2 H'' + \left( \mu + 2 U \right) \left(
c_s^2 - \frac{p}{\rho} - \frac{2}{c_s^2} \left(\rho+p\right)
\frac{d c_s^2}{d \rho} \right) \dot{H} + \left(
\left( 2  c_s^2 -1
\right) \nu + 2  c_s^2W \right) H'  \nn \\
& + & \left\{\left(\rho + p \right) \left[ \left(\rho + p
\right) \left(\mu + 2 U \right)^2 \frac{1}{c_s^2} \left(
\frac{d^2 c_s^2}{d \rho^2}   - \frac{2}{c_s^2}
\frac{d c_s^2}{d \rho}^2 \right)  + \left[ \left(\mu+ 2 U \right)^2
\left( 2+ \frac{\rho- p}{\rho  c_s^2}   \right) \right.  \right. \right. \nn \\
& + & \left. \left. \left.   \frac{1}{c_s^2} \left( 3 U^2 -
\dot{\mu}  - \left(2 \nu + W \right) W + 8 \pi p + \frac{1}{r^2}
\right) \right] \frac{d c_s^2}{d \rho} + 4 \pi
\left(1 + 3 c_s^2 \right) \right]   - \frac{l \left( l+1
\right)}{r^2} c_s^2\right\} H \nn \\
& + & \frac{1}{2} \left( c_s^2 -1 \right) \nu \left( \chi' -
k' \right) + c_s^2  \mu  \dot{\chi} + \frac{c_s^2}{2} \left(
\mu + 2  U \right) \left( 1 + c_s^2 - \frac{p}{\rho} \right) \
\dot{k} + \left\{c_s^2  \left[ 2 \left( 2 \nu + W \right) W + 4
\pi \left( \rho - p \right) \right. \right.  \nn \\
& + & \left.  \left.    2 \dot{\mu} - \frac{2}{r^2} + \left( 1
+ \frac{p}{\rho} -  c_s^2 \right) \mu^2 - 2 \left( 1 + c_s^2 -
\frac{p}{\rho} \right)  \mu  U - 2 U^2  \right] -2 \nu^2 \right\}
\left( \chi + k \right)  \nn \\
& + & \frac{c_s^2}{2} \left( \left(1 +c_s^2 \right) \mu - 2
\left( 1 -c_s^2\right) U \right)  \psi' +   \left\{
\frac{1}{2} \left(1 +c_s^2 \right) \left( c_s^2 \mu' +
\dot{\nu} \right) - \frac{1}{2} \left( \rho + p\right)
\frac{d c_s^2}{d \rho}  \left( 1 - \frac{1}{c_s^2}
\right)
\left(\mu + 2 U\right) \nu  \right. \nn \\
& + & \left. \left[  c_s^2 \left( 1 - \frac{2  p}{\rho} +3
c_s^2  \right) W + \left[ c_s^2 \left( c_s^2 + \frac{p}{\rho} -3
\right) + \frac{p}{\rho}  \right] \nu \right]  U +
\left[  c_s^2 \left( 1 - \frac{p}{\rho} + 3 c_s^2  \right) W \right. \right. \nn \\
& + & \left.  \left. \frac{1}{2} \left[ c_s^2 \left( c_s^2 - 2
\right) + \frac{p}{\rho} \left( 1 + c_s^2 \right) -3 \right] \nu
\right]  \mu \right\} \psi + c_s^2 \left[ 2 \mu - \frac{p}{\rho}
\left( \mu + U \right) \right] \ga' +
\left\{\left( 1 -c_s^2 \right)  \left( c_s^2 \mu' + \dot{\nu} \right)
\right. \nn \\
& + & \left. \left(\rho + p \right) \left( 1 +  \frac{1}{c_s^2}
\right) \left( \mu + 2 U \right) \nu \frac{d
c_s^2}{d \rho} + \left[  2   c_s^2   \left( 1 + c_s^2 - 2
 \frac{p}{\rho} \right) W - 2 \left[ c_s^2 \left( 1 + c_s^2
+ \frac{p}{\rho} \right)  - \frac{p}{\rho}  \right] \nu   \right] U \right. \nn \\
& + & \left. \left[  2   c_s^2   \left( 1 - c_s^2 -
\frac{p}{\rho} \right) W +  \left[ c_s^2 \left( 2 - c_s^2 \right)
+ \frac{p}{\rho} \left( 1-c_s^2 \right)  -1   \right] \nu \right]  \mu
\right\} \ga  \nn \\
& + &  \left\{\left[ 8 \pi \left(\rho + p \right)  c_s^4 +
\left( \frac{l\left( l+1 \right)}{r^2} \frac{p}{\rho} + 8 \pi
\frac{\rho^2 - p^2}{\rho}  \right)   c_s^2 \right] \left( \mu +
2 U \right)  - 2  \frac{l\left( l+1 \right)}{r^3}  c_s^2 \
\dot{r}  \right\} \alpha = 0 \,.
\end{eqnarray}

\section{Gauge invariance} \label{AppGI}

In this Section we show the gauge-invariant character of the perturbative
quantities (\ref{pk}-\ref{pt2}) and (\ref{algi}-\ref{omgi}) at the
perturbative order $(1,1)$.   To that end we follow the procedure
described in Section~\ref{GI11}.  Namely, we
determine, for each quantity, the corresponding term $\sum_\sigma
{\cal I}^{(1,0)}_\sigma   {\cal J}^{(0,1)}_\sigma $ in the expansion
(\ref{g11}) and the gauge transformation of ${\cal H}^{(1,1)}$
[see Eq.~(\ref{ha11_B})]. Then, considering the gauge transformations for the
non-radial perturbations ${\cal J}^{(0,1)}$ and introducing all the
terms in (\ref{agi11_red}), we will prove the gauge invariance
of our perturbations.

The non-radial gauge transformations (which we will use later) are
\be
\tilde p^{(0,1)}_{A} = p^{(0,1)}_{A}+ \hat{\xi}_{A}\,,~~~~
\tilde{G}^{(0,1)} = {G}^{(0,1)} + 2\xi.  \label{gt11}
\ee

\emph{Metric perturbations.}
In Section~\ref{coupsec} we have shown the gauge invariance of the
metric perturbation $k_{AB}^{(1,1)}$. Here we prove that of the metric
perturbation $k^{(1,1)}$. From (\ref{Lieg10}) we find that,
\beq
& & \sum_\sigma {\cal I}^{(1,0)}_\sigma
{\cal J}^{(0,1)}_\sigma  =  2\met^{(1,0) A B}  v_{B} p^{(0,1)}_{A}\,,
\label{expk} \\
& & \tilde{\cal H}^{(1,1)} = {\cal H}^{(1,1)} - 2v^{A}\met_{AC}^{(1,0)}
\xi^{C} \,,
\eeq
where $v_{A} = r_{\mid A}/r\,.$
Therefore using (\ref{gt11}) we find from (\ref{agi11_red})
that $\tilde{k}^{(1,1)} = k^{(1,1)}\,.$ \\

\noindent \emph{Stress-energy tensor perturbations.} They are described
by the seven quantities (\ref{pTAB}-\ref{pt2}). The corresponding
terms  $\sum_\sigma {\cal I}^{(1,0)}_\sigma {\cal J}^{(0,1)}_\sigma$ are:

\begin{eqnarray}
T_{AB}  & :  \qquad &  p^{(0,1)C}    t^{(1,0)}_{A B \mid C} +
t^{(1,0)}_{AC}  p^{(0,1)  C}_{\mid B} + t^{(1,0)}_{CB}
 p^{(0,1)  C}_{\mid A}      - \bar{t}_{A B \mid C}
 \met^{(1,0) C D}  p^{(0,1)}_D    \nonumber \\
& & {} - \bar{t}_{AC} \left( \met^{(1,0)  C D}_{\mid B} \
p^{(0,1)}_D + \met^{(1,0)  C D}  p^{(0,1)}_{D  \mid B}
\right)
- \bar{t}_{BC}  \left( \met^{(1,0)  C D}_{\mid A}
 p^{(0,1)}_D + \met^{(1,0)  C D}  p^{(0,1)}_{D  \mid A} \right)\,, \label{extab} \\
T^3 & : \qquad &  \left( \bar{Q}_{\mid A} +  2 \bar{Q}   v_A
\right) \met^{(1,0) AB}  p^{(0,1)}_{B}   - \left( Q^{(1,0)}_{\mid A}
+ 2 Q^{(1,0)} v_A \right)\bar{\met}^{AB} p^{(0,1)}_{B} +
\frac{l(l+1)}{2}   Q^{(1,0)}   G^{(0,1)}\,, \label{ext3} \\
T_A & : \qquad  &  \left( t^{(1,0)  C}_A - \bar{t}_{AB}
\met^{(1,0) BC} \right) p^{(0,1)}_{C} - \frac{r^2}{2}
Q^{(1,0)} G^{(0,1)}_{\mid A}   \,,   \label{exta} \\
T^2 & : \qquad &  r^2   Q^{(1,0)}   G^{(0,1)}   \,.\label{ext2}
\end{eqnarray}
where $Q$ is the pressure. The gauge transformation of the
${\cal H}^{(1,1)}$ part of the quantities under discussion is given by
$\pounds_{\xi (0,1)}  t_{\alpha\beta}^{(1,0)}$, where  the energy-momentum
tensor for radial perturbations has a block diagonal form
\be
t_{\alpha\beta}^{(1,0)} = \textrm{diag} \left( t_{AB}^{(1,0)} ;
 r^2 Q^{(1,0)} \gamma_{ab} \right)  \,. \label{tAB10}
\ee
Then, we find
\beq
\pounds_{\xi_{(0,1)}} t_{AB}^{(1,0)} & = &
 \pounds_{\hat\xi}  t_{AB}^{(1,0)} \,, \\
\pounds_{\xi_{(0,1)}}g_{Aa}^{(1,0)} & = &   \left( t_{AC}^{(1,0)}
 \hat{\xi}^C + r^2  Q^{(1,0)} \xi_{\mid A} \right)   Y_a \,,\\
\pounds_{\xi_{(0,1)}}   t^{(1,0)}_{ab} & = & r^2 \left(
Q^{(1,0)}_{\mid C}  \hat{\xi}^C - l(l+1)  Q^{(1,0)}  \xi  +
2 v_C  Q^{(1,0)} \hat{\xi}^C \right)   Y   \gamma_{a  b}
+\left(2 r^2 Q^{(1,0)}\xi  \right) Z_{ab}  \,.
\eeq
Therefore the gauge transformations of the $ {\cal H}^{(1,1)} $ terms
for the above quantities are:
\beq
T_{AB} & : \qquad &  \hat{\xi}^C  t^{(1,0)}_{A B \mid C}
+  t^{(1,0)}_{CB}  \hat{\xi}^C_{\mid A} + t^{(1,0)}_{AC}
\hat{\xi}^C_{\mid B}  -  \bar{t}_{A B \mid C}  \met^{(1,0)C}_D
\hat{\xi}^D - \bar{t}^C{}_A  \left( \met^{(1,0)}_{C D \mid B}
\hat{\xi}^D + \met^{(1,0)}_{C D}  \hat{\xi}^D_{\mid B} \right) \nn  \\
&& {} - \bar{t}^C{}_B  \left( \met^{(1,0)}_{C D \mid A} \
\hat{\xi}^D + \met^{(1,0)}_{C D}  \hat{\xi}^D_{\mid A} \right) \,,
\label{gttab} \\
T^3 & : \qquad & \left( Q^{(1,0)}_{\mid A} + 2 Q^{(1,0)} v_A \right)
\hat{\xi}^A - l(l+1)   Q^{(1,0)}  \xi  - \left( \bar{Q}_{\mid A}
+ 2\bar{Q}  v_A \right)  \met^{(1,0) A}_B \hat{\xi}^{B}\,, \label{gtt3}  \\
T_A & : \qquad & t^{(1,0)}_{AB}  \hat{\xi}^B + r^2 Q^{(1,0)}
\hat{\xi}_{\mid A} - \bar{t}_{AB}  \met^{(1,0) BC} p_{C}^{(0,1)}\,,
\label{gtta} \\
T^2 & : \qquad &  2  r^2  Q^{(1,0)}  \xi   \,. \label{gtt2}
\eeq
Combining all this terms into the relation
(\ref{agi11_red}), we finish  the proof of the gauge invariance of
(\ref{pTAB}-\ref{pt2}) at the perturbative order $(1,1)$. \\

\noindent The fluid perturbations for a barotropic fluid
that have to be considered are the two components of the velocity
(\ref{algi}-\ref{gamgi}) and the energy density (\ref{omgi}).
Applying the same procedure we find the corresponding
 $\sum_\sigma {\cal I}^{(1,0)}_\sigma   {\cal} J^{(0,1)}_\sigma$ terms
\begin{eqnarray}
\alpha & : \quad &   \met^{(1,0)}_{A B}  p^{(0,1)  A}
 \bar{u}^B -  p^{(0,1)}_{A}  u^{(1,0)  A} \,, \label{exal} \\
\gamma & : \quad &   \left( n^{(1,0)  A} -
\met^{(1,0)  AC}  \bar{n}_C \right) \left( \delta u^{(0,1)}_A
- \frac{1}{2}  h^{(0,1)}_{AB}  \bar{u}^B  - \bar{u}_{A\mid B}
 p^{(0,1)  B} - p^{(0,1)}_{[B \mid A]}  \bar{u}^B\right)  \nn \\
& & {} - \bar{n}^A \left[ \left( u^{(1,0)  B} - \met^{(1,0)BC}
\bar{u}_C \right)  \left( p^{(0,1)}_{[B \mid A]}
+ \frac{1}{2}  h^{(0,1)}_{AB} \right) +
\left( u^{(1,0)}_{A \mid B} - \Gamma^{(1,0)  D}_{AB}
\bar{u}_{D} \right)  p^{(0,1)  B} \right.  \nn \\
& &  \left. - \met^{(1,0) BD}  \bar{u}_{A \mid B}  p^{(0,1)}_D \right] \,,
\label{exga}  \\
\omega & : \quad &  \met^{(1,0)  A B}  p^{(0,1)}_B
 \bar\Omega_{\mid A} - p^{(0,1)  A} \Omega^{(1,0)}_{\mid A}\,.
\label{exom}
\end{eqnarray}
The gauge transformations (\ref{ha11_B}) for ${\cal H}^{(1,1)}$
are now determined by the Lie derivative of the radial perturbations
of the fluid velocity and energy density
\begin{equation}
u^{(1,0)}_{\alpha} \equiv  \left( u^{(1,0)}_{A},  0 \right)\,,~~~~
\Omega^{(1,0)} \equiv \Omega^{(1,0)} (t,r)\,,
\end{equation}
where $\Omega \equiv \ln \rho $. We obtain
\beq
\pounds_{\xi (0,1)}  u_A^{(1,0)} & = & \hat{\xi}^B
 u_{A \mid B}^{(1,0)} + u_B^{(1,0)}  \hat{\xi}^B_{\mid A} \,,  \\
\pounds_{\xi (0,1)} u_{a}^{(1,0)} & = & u_A^{(1,0)}  \hat{\xi}^A Y_a \,, \\
\pounds_{\xi (0,1)}  \Omega^{(1,0)} & = &\hat{\xi}^A \Omega_{\mid A} \,.
\eeq
The corresponding ${\cal H}^{(1,1)}$ transformations are
\beq
\alpha & : \quad &  u^{(1,0)}_A   \hat{\xi}^{A} -
\met^{(1,0)}_{A B}\hat{\xi}^{A}  \bar{u}^B \,, \label{gtal}  \\
\gamma & : \quad &   \bar{n}^A \left\{\hat{\xi}^{B} \
u^{(1,0)}_{A \mid B} + \left( u^{(1,0)  B} -  \met^{(1,0)  B C}
 \bar{u}_C \right) \hat{\xi}_{B \mid A} - \met^{(1,0)  B C}
 \bar{u}_{A \mid B}  \hat{\xi}_{C} - \Gamma^{(1,0)  B}_{AC}
 \hat{\xi}^{C}  \bar{u}_B \right\} \,, \label{gtga}  \\
\omega & : \quad & \hat{\xi}^A  \Omega^{(1,0)}_{\mid A} -
\met^{(1,0)  A B}   \hat{\xi}_A  \bar\Omega_B \,, \label{gtom}
\eeq
It is again easy to verify the gauge invariance of these perturbations
by bringing all terms into the relation (\ref{agi11_red}).
The fluid perturbation $H^{(1)}$, defined in Eq.~(\ref{En_11}), can be
expanded like the previous quantities,
\beq
H^{(0,1)} & = & {\cal H}^{(0,1)} = \frac{\bar{c}_s^2
\bar{\rho}}{\bar \rho + \bar p}\omega^{(0,1)} \,, \\
H^{(1,1)} & = & {\cal H}^{(1,1)} + \sum_\sigma {\cal I}^{(1,0)}_\sigma
{\cal} J^{(0,1)}_\sigma \,,
\eeq
where
\beq
{\cal H}^{(1,1)} & = & \frac{\bar{c}_s^2   \bar{\rho}}{\bar \rho + \bar p}
\omega^{(1,1)} \,, \\
{\cal I}^{(1,0)}_\sigma & = & \left[ \bar c_s^2 + \bar \rho
\left( \frac{d\bar c_s^2}{d\bar\rho} - \left(1 + \bar{c}_s^2 \right)
\frac{\bar{c}_s^2}{\bar \rho + \bar p}\right)\right]
\frac{\bar{\rho}}{\bar \rho + \bar p}     \omega^{(1,0)} \,, \\
J^{(0,1)} & = & \omega^{(0,1)} \,.
\eeq
Therefore, the gauge-invariant character of $H^{(0,1)}$ and $H^{(1,1)}$, having
fixed the gauge for radial perturbations, follows from the gauge invariance of
$\omega^{(0,1)}$ and $\omega^{(1,1)}$, which has already been proved
previously.

\section{Connection to the Regge-Wheeler metric variables} \label{AppRW}
The perturbations of spherical stars and Black Holes are commonly 
studied in the Regge-Wheeler (RW) gauge \cite{Regge:1957}. Therefore, in this Appendix
 we provide the relations  between the perturbative
variables used in this work and those of RW.
In the RW gauge the linear polar non-radial perturbations assume the following expansion
in spherical tensor harmonics,
\begin{eqnarray}
\left . g_{\alpha\beta} ^{(0,1)} \right|_{RW} & = & \left(\begin{array}{cccc}
 H_{0 , \,  lm}^{(0,1)} \, e^{2\, \Phi } &  H_{1 , \,  lm}^{(0,1)}  &  0 & 0 \\
\\
 H_{1  ,\,  lm}^{(0,1)}  &   H_{2 , \,  lm}^{(0,1)} \, e^{2\, \Lambda } & 0& 0   \\
\\ 0 & 0 &  r^2   K_{lm}^{(0,1)}  & 0\\ \\ 0 & 0 &  0 & r^2 K_{lm}^{(0,1)} \sin ^2 \!  \theta
\end{array}\right)  \,  Y^{lm} \, ,
\end{eqnarray}
where $ H_{0,  \,  lm}^{(0,1)}$, $ H_{1, \,  lm}^{(0,1)} $, $ H_{2, \,  lm}^{(0,1)}$,
$ K^{(0,1)}_{lm}$ are functions of
$(t,r)$.  The Einstein equations for a spherical star imply that
$ H_{0 , \,  lm}^{(0,1)} = H_{2 , \, lm}^{(0,1)} $.
We can also choose the RW gauge for the perturbative variables describing the
coupling by enforcing the RW form of the polar metric perturbations at the $(1,1)$ perturbative
order.  Then, we impose the following form of 
$g_{\alpha\beta} ^{(1,1)}$:
\begin{eqnarray}
\left . g_{\alpha\beta} ^{(1,1)} \right|_{RW} & = & \left(\begin{array}{cccc}
 H_{0, \,  lm}^{(1,1)} \, e^{2\, \Phi } &  H_{1, \,  lm}^{(1,1)}  &  0 & 0 \\
\\
 H_{1, \,  lm}^{(1,1)}  &   H_{2, \,  lm}^{(1,1)} \, e^{2\, \Lambda } & 0& 0   \\
\\ 0 & 0 &  r^2   K_{lm}^{(1,1)}  & 0\\ \\ 0 & 0 &  0 & r^2 K_{lm}^{(1,1)} \sin ^2 \! \theta
\end{array}\right)  \, Y^{lm} \, ,
\end{eqnarray}
where also $ H_{0, \,  lm}^{(1,1)}$, $ H_{1, \,  lm}^{(1,1)}
$, $ H_{2, \,  lm}^{(1,1)}$, $ K^{(1,1)}_{lm}$ are functions of
$(t,r)$.

The expansion of the linear non-radial and the  coupling metric
perturbations in terms of the GSGM variables can be derived from (\ref{newlab})
by applying the 2-parameter expansion of GSGM formalism described in section \ref{coupsec}.
First of all, we must first apply the RW gauge at first and
second perturbation order, i.e, $$ h_{A, \, lm} ^{(0,1)} = G_{lm}
^{(0,1)} = h_{A, \, lm} ^{(1,1)} = G_{lm} ^{(1,1)}  = 0\,.$$ 
Then, we take the perturbative expansion of the gauge-invariant tensor $k_{AB}$ 
(\ref{KABframe}), written in the basis of $M^2$ spanned by the vectors (\ref{urdbg},\ref{nrdbg}),
and use the definition (\ref{chidef}) for the $\chi$ perturbation. 
Finally, taking into account Einstein's equation (\ref{eta}), we find 
the following relations: (i) For the $(0,1)$ linear perturbations, 
\be 
H_{0 \, ,lm}^{(0,1)} = H_{2 \, , lm}^{(0,1)} = \chi_{lm}^{(0,1)} +
k_{lm}^{(0,1)} \, ,\qquad H_{1 \, , lm}^{(0,1)} = - \psi
_{lm}^{(0,1)} \,e^{ \Phi + \Lambda } \, , \qquad K^{(0,1)}_{lm} =
k^{(0,1)}_{lm}    \,. \ee
(ii) For the $(1,1)$  coupling perturbations,
\begin{eqnarray}
H_{0 \, , lm}^{(1,1)}  & = &   \chi^{(1,1)}_{lm} + k^{(1,1)} _{lm}  +
\left( 2 \, \eta ^{(1,0)} - \chi ^{(1,0)}\right) \left( \chi^{(0,1)}_{lm} + k^{(0,1)} _{lm} \right) + 2 \,
\ga ^{(1,0)} \psi ^{(0,1)}_{lm}  \, ,
    \\
H_{1 \, , lm}^{(1,1)}  & = & - \left[  \psi^{(1,1)}_{lm}
+ 2 \, \ga ^{(1,0)} \, \left( \chi^{(0,1)}_{lm} + k^{(0,1)} _{lm} \right)  + \eta ^{(1,0)} \psi ^{(0,1)}_{lm}
\right] \,e^{ \Phi + \Lambda }   \, , \\
H_{2 \, , lm}^{(1,1)} & = & \chi^{(1,1)}_{lm} + k^{(1,1)} _{lm} 
+  \chi^{(1,0)}  \left( \chi^{(0,1)}_{lm} + k^{(0,1)} _{lm} \right)
+ 2 \, \ga ^{(1,0)} \psi ^{(0,1)}_{lm}   \, ,\\
K^{(1,1)}_{lm}  & = & k^{(1,1)}_{lm}  \, .
\end{eqnarray}
Before concluding this section we also give here the relations even between
the quantities used in this paper and those considered by
Allen et al. \cite{Allen:1998xj} and Ruoff \cite{Ruoff:2001ux} at
linear order:
\begin{equation}
 \chi ^{(0,1)} = \left\{ \begin{array}{ll}
r \, e^{-2\, \Phi} \, S  & \mbox{~~in~\cite{Allen:1998xj}\,,} \\
r \, S & \mbox{~~in~\cite{Ruoff:2001ux}\,,}
\end{array} \right.
\qquad \qquad 
k ^{(0,1)}= \left\{ \begin{array}{ll}
F/r & \mbox{~~in~\cite{Allen:1998xj}\,,} \\
T/r & \mbox{~~in~\cite{Ruoff:2001ux}\,.}
\end{array} \right.
\end{equation}
For more details about the connection between the GSGM and other formalisms see the appendix
in \cite{Gundlach:1999bt}.




\end{document}